\documentclass{ws-rv9x6}
\usepackage{ws-rv-thm}   % comment this line when `amsthm / theorem / ntheorem` package is used
\usepackage{ws-rv-har}   % harvard style, author-date 
\usepackage[hidelinks]{hyperref}
\makeindex
\usepackage{graphicx}	% Including figure files
\usepackage{amsmath}	% Advanced maths commands (including split to split equations)
\usepackage{aas_macros} % journals

\newcommand{\Qd}{Q_\mathrm{D}^{\star}}

\begin{document}

\chapter[Planetesimal/debris discs]{Planetesimal/Debris discs}

\author[S. Marino]{Sebastian Marino}% \footnote{}}
%\index[aindx]{Author, F.} % or \aindx{Author, F.}
%\index[aindx]{Author, S.} % or \aindx{Author, S.}

\address{Jesus College, University of Cambridge\\
Jesus Lane, Cambridge, CB5 8BL, UK \\
sebastian.marino.estay@gmail.com} % \footnote{Affiliation footnote.}

\begin{abstract}
  This review chapter for young researchers presents our current
  understanding of debris discs. It introduces some of their basic
  properties and observables, and describes how we think they form and
  collisionally evolve. Special emphasis is dedicated to ALMA
  observations of the dust and gas, which constrains the distribution
  of planetesimals, their volatile composition, and potential volatile
  delivery to planetary atmospheres.
\end{abstract}

%\markboth{Even Page Header}{Odd Page Header} % Customized running heads

\body

%\tableofcontents

\section{What are debris discs?}

 As usual, establishing an astrophysical definition for a broad
 phenomenon that sometimes overlays with other types of circumstellar
 discs is challenging. Hence, it is perhaps clearer to answer this
 from a Solar System perspective. The Solar System's debris disc is
 made of all its non-planetary components. Namely, the Asteroid belt,
 the Kuiper belt, the Oort cloud, the Zodiacal cloud, Jupiter family
 comets, etc. These are a natural part of planetary systems, and today
 we know of hundreds of extrasolar analogues.

The study of the Solar System's debris disc has been fundamental for
the development of planetary dynamics and elucidating the and history
of the Solar System planets. Similarly, studying debris discs around
other stars is crucial for obtaining a holistic view of planetary
systems. Most debris discs are discovered thanks to the infrared (IR)
excess emitted by circumstellar dust, first detected around a few
nearby stars almost four decades ago \citep{Aumann1984}. The fact that
dust is short-lived against collisions and radiation forces compared
to the ages of these systems \citep{Backman1993}, means that the
observed dust is continuously replenished by collisions of
longer-lived bodies such as km-sized asteroids or comets. Since then,
multiwavelength studies have shown that these dusty discs are a
ubiquitous component of planetary systems, they also possess a gas
component, and our Solar System's debris disc might not the norm. Dust
and gas observations allow us to study this non-planetary component
and constrain their composition, evolution and dynamics.

This chapter aims to serve as an introduction to debris discs and
summarise the latest advances in this field, making a special emphasis
in what we have learnt from observations of gas and dust with the
Atacama Large Millimeter/submillimeter Array (ALMA) in the last 3
years. Previous reviews by \citet{Wyatt2008}, \citet{Krivov2010},
\citet{Matthews2014pp6} and \citet{Hughes2018} cover at great length
the initial discovery of debris discs around other stars since the
launch of the Infrared Astronomical Satellite (IRAS), their basic
observables, detection limits, their expected evolution, the diversity
of disc structures, dust properties, and variability, and thus are
excellent references for eager readers. One important class of debris
discs that is not covered here is the discs around polluted white
dwarfs, which has been recently covered in detail by
\citet{Veras2021}.

\section{Basic properties and observables}
\label{sec:basic}
The solid component of debris discs is made of solids with a wide size
distribution, ranging from $\mu$m-sized dust up to km-sized or even
larger planetesimals. This distribution is regulated by collisions
that grind down solids moving mass from large to small bodies, and the
removal of small dust by radiation pressure and Poynting-Robertson
(P-R) drag \citep[e.g.][]{Lohne2008}. Since the expected size ($a$)
distribution is close to $dN\propto a^{-3.5} da$, the bulk of the
cross sectional area is dominated by the smallest grains at the bottom
of the size distribution, while the mass is dominated by the largest
bodies. This means that debris disc observations trace the smallest
debris that are constantly destroyed and replenished in this
collisional cascade.

\begin{figure}[h!]
  \begin{center}
    \includegraphics[width=0.6\textwidth]{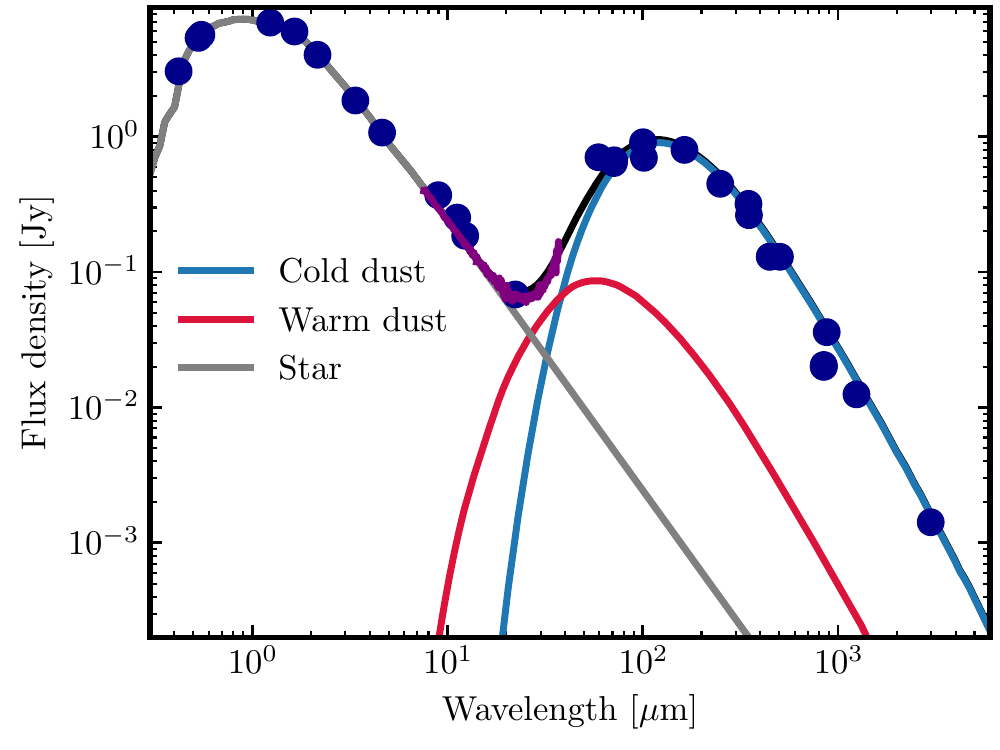}
    \caption{Spectral energy distribution of HD~107146. The blue
      points and purple line represent photometric measurements and an
      Spitzer/IRS spectrum, respectively. The solid lines represent
      the emission from the star (grey) and two dust components with
      different temperatures (adapted from
      \citealt{Marino2018hd107}).} \label{fig:sed}
  \end{center}
\end{figure}

Two fundamental properties of debris discs can be obtained from
unresolved observations that constrain their spectral energy
distribution (SED). First, we can derive the blackbody temperature of
the emission using Wien displacement law. Then, assuming dust behaves
approximately like blackbodies, this temperature ($T_{\rm BB}$) and
the known stellar luminosity ($L_\star$) can be used to infer the
characteristic radius of the disc, sometimes called blackbody radius
$r_{\rm BB}$,
\begin{equation}
  r_{\rm BB} = 1\ {\rm au} \left( \frac{278.3\ {\rm K}}{T_{\rm BB}} \right)^{2}\left( \frac{L_\star}{L_{\odot}} \right)^{0.5}. 
\end{equation}
Note that $r_{\rm BB}$ is usually smaller than the true radius
($r_{\rm belt}$) due to small grains being inefficient emitters at IR
wavelengths, hence having higher equilibrium temperatures than
blackbodies \citep{Booth2013, Kennedy2014, Pawellek2014}. Some systems
can also display more than one temperature, typically with a peak of
emission at far-IR wavelengths and significant emission in the
mid-IR. Figure \ref{fig:sed} shows an example of such a system with
two temperature components. Such systems are inferred to have a cold
Kuiper belt analogue at tens of au, i.e. an \textit{exoKuiper belt},
and a warm component analogous to the Asteroid belt or Zodiacal cloud
closer in \citep[e.g.][]{Backman2009, Chen2009, Morales2009,
  Ballering2014, Kennedy2014}.

The second most important property is the disc fractional luminosity
($f_{\rm IR}$), defined as the ratio between the dust and stellar
luminosities. Since debris discs are optically thin, $f_{\rm IR}$ is
approximately $\sigma_{\rm tot}/(4\pi r_{\rm belt}^2)$, where
$\sigma_{\rm tot}$ is the total absorption cross-sectional area of the
dust. Therefore, the SED provides direct constraints to the amount of
dust in a system.

There are additional constraints that can be derived from SEDs. For
example, the spectral slope of the dust emission depends on the dust
opacity, which depends on the size distribution. Thus, it is in
principle possible to constraint the size distribution and test
collisional models \citep{Ricci2015, MacGregor2016}. However,
uncertainties about the composition and structure of grains means that
size distribution measurements suffer from great systematic
uncertainties \citep{Lohne2020}.

% frequency
Far-IR surveys with \textit{Spitzer} and \textit{Herschel} have
characterised the frequency of cold debris discs around nearby stars
with fractional luminosities $\gtrsim10^{-6}$, obtaining detection
rates of $30\%$ around A stars \citep{Su2006} and about 20\% for FGK
stars \citep{Eiroa2013, Sibthorpe2018}. Note that these percentages
are not occurrences, but rather detection rates above a given
sensitivity limit. As a reference, the Kuiper belt would be more than
an order of magnitude below the detectability threshold of such
surveys, and thus it is possible that every star has a debris disc at
least as massive as the Kuiper belt. This sensitivity limit is
specially poor for surveys around M stars, which have yielded
detection rates of only 2-14\% as these discs tend to be colder and
thus harder to detect in the far-IR \citep{Kennedy2018,
  Luppe2020}. Moreover, the detection rates are also sensitive to the
age of the sample considered. \citet{Pawellek2021} showed how debris
discs are found around $\sim75\%$ of F stars in the 23 Myr old $\beta$
Pic Moving Group. This suggests that the real incidence of massive
debris discs could be much higher, and heavily dependent on the age.

% correlations with planets
Recent studies have also searched for correlations between the
presence of planets and debris discs. \citet{Meshkat2017} looked at
$5-20$~$M_{\rm Jup}$ planets at separations $10-1000$~au, detected
through direct imaging, and found tentative evidence of a higher
occurrence rate of planets in systems with detected debris
discs. Therefore, it appears that the formation of massive exoKuiper
belts is favoured in systems that also form gas giants at wide
separations. On the other hand, \citet{Yelverton2020} found no
significant difference in the fractional luminosity or temperature
distributions of debris discs around stars with and without close-in
planets detected through radial velocities. Finally, it is worth
noting that debris discs detection rates seems to be unaffected by the
presence of binaries at separations greater than $135$~au, halved by
the presence of binaries at separations smaller than $25$~au, and
reduces to zero for binaries with intermediate separation that roughly
overlay with typical debris disc sizes \citep{Yelverton2019}.

\subsection{Debris disc formation}

The existence of debris discs requires the formation of planetesimals
in protoplanetary discs. These km-sized bodies must grow from the dust
in protoplanetary discs. However, their formation encounters multiple
barriers that prevent dust from growing beyond cm sizes due to their
radial drift \citep{Whipple1973, Weidenschilling1977drag}, or bouncing
and fragmentation \citep{Zsom2010, Blum2008}. The first barrier can be
circumvented by a pressure maxima in the gas, which act as traps where
dust can survive, accumulate and efficiently grow to cm sizes. Strong
evidence of such dust traps is seen in the axisymmetric rings
\citep[][]{Dullemond2018dsharp, Rosotti2020} and clumps in
protoplanetary discs \citep[][]{Casassus2019}.

Two mechanisms have been proposed to overcome the second
barrier. First, dust grains could grow as very porous fractals that
remain sticky and well coupled to the gas, bypassing both the radial
drift and fragmentation/bouncing barrier \citep{Kataoka2013}. This
relies, however, on the assumption of a high stickiness which is still
uncertain. A second and perhaps more robust scenario is the streaming
instability \citep{Youdin2005, Johansen2007}. This instability is
triggered when the dust densities reach values similar to the gas
densities, and dust is large enough to be slightly decoupled from the
gas \citep[see][]{Li2021}. This instability results in the collapse of
dust clouds and the formation of $\sim100$~km-sized planetesimals. It
has been shown that dust growth, their settling to the midplane and
concentration in radial or azimuthal dust traps (as the ones observed)
is enough to trigger the streaming instability \citep{Stammler2019,
  Carrera2021}. The newborn planetesimals might continue growing via
mutual collisions, but never reach a mass high enough to clear their
orbits. This is the last condition for debris disc formation, which is
not problematic given the extremely long growth timescales of
planetesimals at tens of au \citep[e.g.][]{Kenyon2008}.

To constrain debris disc formation further, recent ALMA surveys have
looked at class III stars, i.e. those that have recently lost their
protoplanetary discs, to look for the youngest debris discs and
compare them with the older populations \citep{Lovell2021classIII,
  Michel2021}. For example, the debris disc detection rate in class
III stars in Lupus is consistent with population models that fit older
stars, suggesting that debris disc formation does not require
long-lived protoplanetary discs \citep{Lovell2021classIII}.

\nopagebreak

\section{Collisional evolution}
\label{sec:coll}
After the dispersal of protoplanetary discs, planetesimals orbits can
be stirred, for example, through secular interactions with planets
\citep{Mustill2009}, or through close encounters with massive
planetesimals in the disc \citep[][]{Kenyon2008,
  Krivov2018stirring}. Stirring results in orbit crossing and relative
velocities that are high enough to fragment km-sized planetesimals and
trigger a collisional cascade that resupplies the observed dust. The
collisional evolution has been studied in detail both analytically
\citep[e.g.][]{Dominik2003, Wyatt2007hotdust, Lohne2008}, and using
numerical simulations \citep{Krivov2006, Thebault2007,
  Gaspar2012}. Here I summarise some simple equations that can be used
to estimate the collisional timescales of debris discs.

First, let's consider a size distribution of the form $dN=D^{2-3q}dD$,
where $D$ is the diameter of solids and $-q$ is the power-law index of
the mass distribution ($dN=m^{-q}dm$). Generally speaking, for $q$ in
the range between 5/3 and 2, we have that most of the disc mass is in
the objects close to the maximum size $D_{\max}$ and the surface area
is dominated by solids close to the minimum size $D_{\min}$. In debris
discs, away from the edges of the size distribution, we typically
expect $q$ to take values close to 11/6 ($dN=D^{-3.5}dD$). This value
is only truly valid for an infinite collisional cascade where the
strength of solids is independent of size
\citep{Dohnanyi1969}. Nevertheless, the estimated values are not far
off from it \citep[e.g.][]{Norfolk2021}. Very small grains will be
removed from the system due to the effect of radiation pressure (or
stellar winds around low-mass stars). The critical size is called the
blow-out size and corresponds to a size for which the ratio between
the radiation and gravitational force is 1/2. Smaller grains are put
on hyperbolic orbits once released from larger bodies on nearly
circular orbits exiting the system on very short timescales. Assuming
perfect blackbody grains, the blow-out size can be written as
\begin{equation}
D_{\rm bl}= 0.8 \left(\frac{L_\star}{L_\odot} \right)
\left(\frac{M_\star}{M_\odot} \right)^{-1}
\left(\frac{\rho}{2.7\ \mathrm{g\ cm^{-3}}} \right)\ \mathrm{\mu m},
\end{equation}
where $\rho$ is the density of solids. Now, given a minimum size
$D_{\rm bl}$, maximum size $D_{\max}$ and $q=11/6$, the total solid
mass in a disc is given by
\begin{equation}
M_{\rm tot} = 85 \left(\frac{f_{\rm IR}}{10^{-3}} \right) \left(\frac{r}{100\ \mathrm{au}} \right)^{2} \left(\frac{D_{\rm max}}{10\ \mathrm{km}} \right)^{0.5}\left(\frac{D_{\rm bl}}{1\ \mathrm{\mu m}} \right)^{0.5}\ M_{\oplus}. \label{eq:mass}
\end{equation}

The total disc mass will be a function of time as planetesimals grind
down and small dust is ejected from the system. Generally speaking we
have that $dM_{\rm tot}/dt = -M_{\rm tot}/t_c$, where $t_c$ is the
collisional timescale of the largest planetesimal. We can estimate the
collisional timescale if we further assume the solids strength or
disruption threshold is independent of size, $q=11/6$, the dispersion
of eccentricities and inclinations are equal, and that the smallest
body able to fragment the largest planetesimals is much smaller than
$D_{\max}$. Under these assumptions the collisional timescale of the
largest planetesimal becomes \citep{Wyatt2007hotdust}
\begin{equation}
  \begin{split}
t_c= 450\ \left(\frac{r}{100\ \mathrm{au}}\right)^{13/3}\left(\frac{dr}{r}\right)\left(\frac{D_{\max}}{10\ \mathrm{km}}\right)
\left(\frac{\Qd}{330\ \mathrm{J~kg^{-1}}}\right)^{5/6}\left(\frac{e}{0.1}\right)^{-5/3} \\
\left(\frac{M_\star}{M_\odot}\right)^{-4/3}
\left(\frac{M_\mathrm{tot}}{85\ M_\oplus}\right)^{-1}
\ \mathrm{Myr},
\end{split}
\label{eq:tc}
\end{equation}
%% \begin{equation}
%%   M_\mathrm{tot}(t) = \frac{M_\mathrm{tot}(0)}{1+t/t_c(0)}, \label{eq:Mtott}
%% \end{equation}
where $dr$ is the width of the disc, $\Qd$ is the disruption threshold
of planetesimals. Since $t_c$ is inversely proportional to
$M_\mathrm{tot}$, the disc mass will evolve as $M_\mathrm{tot}(t) =
M_\mathrm{tot}(0)/(1+t/t_c(0))$, where $M_\mathrm{tot}(0)$ and
$t_c(0)$ are the initial disc mass and collisional timescale. For
$t\gg t_c(0)$, the largest planetesimals in a disc are in in
collisional equilibrium and the disc mass and fractional luminosity
will decay as $1/t$. Note that $t=0$ here represents the time at which
the disc is stirred, which could be different from the age of the
system. Regardless of the initial disc mass,
$M_\mathrm{tot}<M_\mathrm{tot}(0) t_c(0)/t$, i.e. the maximum disc
mass as a function of time is
\begin{equation}
  \begin{split}
  M^{\max}_{\rm tot} = 85 \ \left(\frac{r}{100\ \mathrm{au}}\right)^{13/3}\left(\frac{dr}{r}\right)\left(\frac{D_{\max}}{10\ \mathrm{km}}\right)
  \left(\frac{\Qd}{330\ \mathrm{J~kg^{-1}}}\right)^{5/6}\left(\frac{e}{0.1}\right)^{-5/3}\\
  \left(\frac{M_\star}{M_\odot}\right)^{-4/3}\left(\frac{t}{450\ \mathrm{Myr}}\right)^{-1}\ M_\oplus. 
  \end{split}
  \label{eq:Mmax}
\end{equation}

Similarly, we can define a maximum fractional luminosity making use of
Equation~\ref{eq:mass} and \ref{eq:Mmax}. We find
\begin{equation}
   \begin{split}
f^{\max}_{\rm IR} = 10^{-3}\ \left(\frac{r}{100\ \mathrm{au}}\right)^{7/3}\left(\frac{dr}{r}\right)\left(\frac{D_{\max}}{10\ \mathrm{km}}\right)^{0.5} \left(\frac{D_{\min}}{1\ \mathrm{\mu m}}\right)^{-0.5} \\
  \left(\frac{\Qd}{330\ \mathrm{J~kg^{-1}}}\right)^{5/6}\left(\frac{e}{0.1}\right)^{-5/3}
  \left(\frac{M_\star}{M_\odot}\right)^{-4/3}\left(\frac{t}{450\ \mathrm{Myr}}\right)^{-1}.
   \end{split}
   \label{eq:fmax}
\end{equation}

The equations above provide very useful and simple ways of estimating
collisional lifetimes and evolution of discs, which in contrast to
more realistic numerical simulations they allow to quickly compute the
evolution of thousands of models. This is particularly useful for
population synthesis models. Based on these equations and infrared
surveys, \citet{Wyatt2007Astars} and \citet{Sibthorpe2018} showed how
population synthesis models can constrain the average values of
$M_\mathrm{tot}(0)$, $D_{\max}$, $\Qd$ and $e$, which determine the
initial fractional luminosity of discs and their collisional
lifetime. Equation~\ref{eq:fmax} also predicts that a disc older than
$t_c$, will have a surface density that scales with radius as
$r^{7/3}$, independently of the initial surface density
\citep{Kennedy2010}. Thus by constraining the slope of the surface
density distribution of a disc, it is possible to test if $t_c(0)$ is
shorter or longer than the age of the system or stirring timescale
\citep{Marino2017etacorvi, Marino201761vir}. Moreover, a disc inner
edge that is sharper than $r^{7/3}$ might indicate that the disc inner
edge was truncated by a planet \citep[e.g.][]{Marino2018hd107,
  Marino2019, Matra2020, Marino2021}. For a recent and detailed
discussion on the collisional evolution of debris discs and their
inferred masses, see \citet{Krivov2021}.

%% These equations, however, assume a power-law size distribution that
%% remains unchanged during the collisional evolution of the disc. To
%% show how this is not always a good approximation for the evolution of
%% dust, Figure~\ref{fig:collevol} shows the evolution of the disc mass
%% calculated using a simple numerical approach that computes the disc
%% evolution assuming the size distribution is in quasi-steady-state
%% \citet{Wyatt2011, Marino201761vir}. The mass-loss rate due to
%% catastrophic collisions in each size bin is balanced by the input from
%% fragmentation of larger bodies in destructive collisions, which inputs
%% mass into the bin. It also takes into account the size-dependent
%% strength of solids. The left panel....

%% \begin{figure}[h!]
%%   \begin{center}
%%     \includegraphics[trim=0. 0. 12.6cm 0., clip, width=0.49\textwidth]{Figure_Mtot_Mdust_vst.pdf}
%%     \includegraphics[width=0.49\textwidth]{MvsD.pdf}
%%     \caption{.} \label{fig:coll}
%%   \end{center}
%% \end{figure}

\subsection{The outliers}

While the model presented above can account for the majority of discs,
two types of discs remain hard to explain with standard collisional
models.

First, 10-30\% of nearby stars present emission in the near- or mid-IR
that indicates the presence of high levels of hot or warm dust
\citep{Kral2017exozodis, Ertel2020, Absil2021}. The warm dust is
typically located at distances of $\sim1$~au, i.e. analogous to the
Zodiacal cloud in the Solar System and thus often referred as
\textit{exozodis}. While in principle, dust can be produced at any
distance from the star if there are planetesimals colliding, in many
systems the fractional luminosities of the hot and warm dust are
orders of magnitude above the maximum levels expected for their ages
and inferred dust radii \citep{Wyatt2007hotdust}. This means that the
dust is likely produced by an alternative mechanism. In the case of
the hot dust, it is located near the sublimation radius and dominated
by small grains \citep{Kirchschlager2017}. Whilst it is expected that
dust produced in an exterior planetesimal belt will migrate inwards
through P-R drag and reach the inner regions \citep[in the absence of
  very massive planets,][]{Bonsor2018}, this dust does not remain for
long enough in the inner regions to explain the near-IR excesses
\citep{vanLieshout2014, Sezestre2019}. \citet{Pearce2020} recently
showed that the gas originating from the sublimation of dust could
push dust outwards, counterbalance the effect of P-R drag, and thus
explain the pile-up of dust near the sublimation radius. Two main
ideas have been proposed to explain the origin of the warm dust. For a
large fraction of systems, P-R drag would be efficient enough to
explain the observed mid-IR levels \citep{Kennedy2015lbti,
  Rigley2020}. However, the warm dust in some systems needs to be
resupplied faster, which could be achieved via exocomets being
transported in via scattering or resonances \citep{Bonsor2012analytic,
  Bonsor2012nbody, Bonsor2013, Bonsor2014, Marboeuf2016, Faramaz2017,
  Marino2018scat}.

Finally, the warm dust in some systems could be a transient phenomena
such as the product of a recent giant collisions \citep{Jackson2012,
  Genda2015, Watt2021}, which are expected in the last stages of
terrestrial planet formation \citep{Chambers2013}. This is especially
favoured in systems with fractional luminosities above 1\%, which are
termed \textit{extreme debris discs} \citep[EDDs][]{Balog2009}, and
systems displaying mid-IR spectroscopic features that indicate the
presence of very small crystalline dust particles \citep{Rhee2008,
  Lisse2009, Olofsson2012}. Some systems even display variability in
their mid-IR emission as expected in a giant collision scenario
\citep{Meng2014, Meng2015, Su2019, Moor2021, Rieke2021}, or even have
short-lived gas \citep{Marino2017etacorvi, Schneiderman2021}.

%% analytic expressions

%% blow out size

%% numerical sim of size distribution

% outliers, exozodis, hot dust, (comet hypothesis) and gas drag for hot dust

% extreme debris discs and terrestrial planet formation (Moor+2021)

%% The ubiquity
%% of cold debris discs indicates that planetesimal formation at tens of
%% au must be common

%% there the has been significant advancement in in
%% planetesimal formation seems that both barriers

%% Once planetesimals are formed, the last condition for the formation of
%% a debris disc is the

%% Moreover, it photoevaporation

%% occurr in protoplanetary discs , dust growth and their settling to the
%% midplane and concentration in dust traps

%% the dust-rich rings and
%% clumps seen in in protoplanetary discs with ALMA indicate that
%% planetesimal formation could be udergoing

%% conditions to trigger

%% relative to the in which
%% ...

%% Planetesimal
%% formation is

%% The formation of debris discs

\section{Resolved observations}

Whilst unresolved observations provide information about some of the
most fundamental properties of debris discs, it is only through
resolved observations that we can access their detailed
structure. Given the wide size distribution of dust and the grain size
dependence on the opacity, observations at different wavelength trace
different ranges of grain sizes.

\subsection{Scattered light observations}

Observations at short wavelengths are sensitive to smaller grains,
whose distribution is highly affected by radiation pressure or stellar
winds. Both effects can put grains smaller than the blow-out size on
hyperbolic orbits and make slightly larger grains to spiral inwards
towards the star \citep[e.g.][]{Thebault2008, Plavchan2005}. This is
why observations at these wavelengths usually display large halos
beyond the planetesimal disc \citep[e.g. out to 1000~au around
  $\beta$~Pic,][]{Janson2021}. At optical and near-IR wavelengths, the
combination of negligible cold dust thermal emission (see
Figure~\ref{fig:sed}) and high stellar fluxes means that the disc flux
is dominated by stellar light that is scattered by small dust. These
observations are thus often called scattered light observations. In
the last three decades, dozens of scattered light images have been
obtained from the ground and space \citep[e.g][]{Esposito2020,
  Schneider2014}.

Several important properties can be derived from scattered light
observations that relate to the intrinsic structure and composition of
dust grains. First, by measuring the disc scattered light flux
($F_{\rm scat, \lambda}$), fractional luminosity, and stellar flux
($F_{\star, \lambda}$), a first order estimate of the albedo can be
obtained as $F_{\rm scat, \lambda}/(F_{\rm scat, \lambda}+f_{\rm
  IR}F_{\star, \lambda})$ \citep{Marshall2018}. These estimates,
however, typically assume isotropic scattering, while in reality
scattering is anisotropic, making discs brighter towards their near
side \citep[i.e. light is preferentially scattered forward,
  e.g.][]{Kalas2005}. A more robust approach is to characterize how
the surface brightness varies as a function of scattering angle,
i.e. the scattering phase function. Observations of a disc inclined by
$i$ relative to a pole on orientation cover a range of scattering
angles from ($90^{\circ}-i$ to $90^{\circ}+i$). Therefore, highly
inclined discs are ideal to sample this phase function over a wide
range of scattering angles \citep[e.g.][]{Milli2017}. In addition,
polarized scattered light observations can be used to constrain the
degree of polarization \citep{Perrin2015}. Scattered light emission
can also be characterised as a function of wavelength adding more
constraints to the size distribution and composition of grains,
although strong degeneracies still remain \citep{Kohler2008}.

These properties have been characterised for multiple systems finding
some similarities. For example, their phase functions are very similar
showing a strong forward scattering peak, a moderate backward
scattering peak, and a flat phase function at intermediate scattering
angles \citep{Hughes2018}, which are consistent with porous grains
with properties similar to Zodiacal and cometary dust
\citep{Graham2007, Ahmic2009}. Polarised observations of HR~4796 also
suggests grains are porous \citep{Milli2019}, although some studies
have found compact grains are preferred when scattered light and
thermal emission observations are fitted simultaneously
\citep{Rodigas2015, Ballering2016}. On the other hand, albedo
estimates typically range between 0.06-0.6 and typically exclude pure
water ice compositions \citep{Schneider2006, Choquet2018,
  Marshall2018, Romero2021}. This range of albedos in addition to
small, but yet significant, differences in the phase function of discs
suggests a diversity in the properties of small dust in debris
discs. In order to improve these constraints, it is necessary to move
towards comprehensive models that can reproduce multiwavelength
observations, taking into account the dynamics and spatial segregation
of different grains sizes as well as their optical properties
\citep[e.g.][]{Pawellek201949ceti}.

%% Esposito20: Differences in the grain properties of discs around similar stars.
%% Hughes18: phase functions are roughly similar between different systems and zodiacal dust + comets, but there are differences in some discs e.g. in HR4796
%% Graham2007, Ahmic2009:phase function roughly consistent with zodiacal dust properties
%% Grynko2004: generic shape of PF suggests high porosity

%% Milli19: HR4796 polarised phase function incompatible with compact spheres (Mie theory)--> favour agreagete particles.
%% Graham2007: high porosity for AU Mic based on degree of optical polarisation.
%% Ballering2016: compact grains are preferred.
%% Rodigas2015: low porosity (compact grains) are preferred.

%% Debes2008: red - NIR colors of HR4796 suggest organic compounds.
%% Kohler2008: similar behavious can be obtained by porous grins made of solicates, am carbon and water ice.
%% Rodigas2015: beyond 3um colos are neutral and consistent with most species, even water ice fractions up to 50%.

%% (albedo, phase function, and
%% polarization) are commonly used to determine properties of dust grains
%% such as their characteristic size and internal structure, which
%% indicate..... \citep{Milli2019}.

%% In
%% addition, 

\subsection{Thermal emission at mm wavelengths}

\begin{figure}[h!]
  \begin{center}
    \includegraphics[width=1.\textwidth]{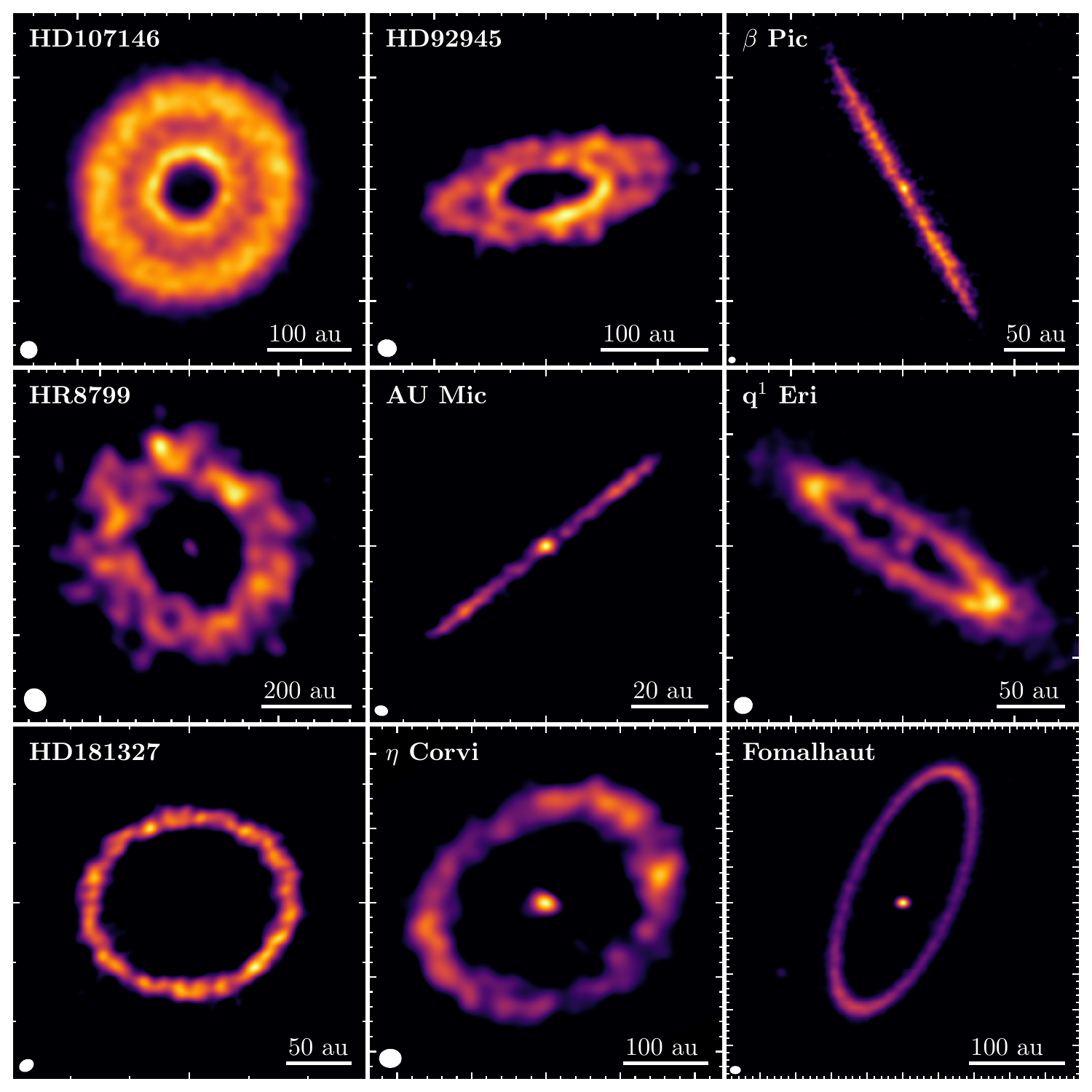}
    \caption{ALMA clean images of 9 exoKuiper belts
      \citep{Marino2018hd107, Marino2019, Matra2019betapic,
        Faramaz2021, Daley2019, Lovell2021q1eri, Pawellek2021,
        Marino2017etacorvi, MacGregor2017}. The white ellipses at the
      bottom left show the beam size, while the white line at the
      bottom right provides a scale in au. The big and small ticks
      along the edges are separated by 5'' and 1'',
      respectively.} \label{fig:continuum}
  \end{center}
\end{figure}

At longer wavelength, the emission is dominated by thermal
radiation. In what follows I will focus on mm ALMA observations. Since
the dust opacity decreases steeply at wavelengths longer than the
grain size, mm observations are dominated by $\sim0.1-10$~mm-sized
grains , which are not affected by radiation pressure. Therefore,
their distribution traces the distribution of their parent
planetesimals. Figure~\ref{fig:continuum} shows a gallery of exoKuiper
belts imaged with ALMA, displaying a diversity of morphologies from
narrow (bottom) to wide belts (middle), some split by gaps (top left
and middle), and one eccentric belt that is offset from the star
(bottom right).

In the following subsections, I provide an overview of the main
features that have been observed in exoKuiper belts using ALMA dust
continuum observations. I start by describing some general trends and
characteristics, such as the radius and width of exoKuiper belts
(\S\ref{sec:mmlaw} and \S\ref{sec:widths}), and then discuss more
detailed substructures that have been found, some of which are
displayed in Figure~\ref{fig:continuum}.

%% \subsection{Our own Solar System}

\subsubsection{Radius distribution}
\label{sec:mmlaw}

One of the most general and first goals of ALMA for debris discs
studies was to resolve a large sample of exoKuiper belts and measure
their radii. As described in \S\ref{sec:basic}, this cannot be
accurately done by looking at SEDs and thus resolved observations are
necessary. ALMA still is the best instrument to image exoKuiper belts
thanks to its high sensitivity and adjustable resolution, which allows
to tailor every observation depending on system properties such as its
distance. During the first years of ALMA, big efforts where put on
imaging every known belt with sufficient flux at mm wavelengths to be
resolved. By 2018, we had resolved 26 exoKuiper belts around AFGKM
stars, which \citet{Matra2018mmlaw} analysed to show that belts tend
to be larger around more massive and luminous stars. This result was
tested against observational biases that would make it hard to detect
cold and large belts around less luminous stars, and collisional
evolution which would quickly deplete small belts around more luminous
stars. Those tests showed that both effects were unlikely to explain
the trend, suggesting that this was a correlation present since
exoKuiper belts are born. It is possible that this relation arises
from planetesimal belts forming at preferred locations in
protoplanetary disks (e.g. near the CO snow line), which scale with
stellar properties, but this hypothesis has not been confirmed by
models or further observations.

\begin{figure}[h!]
  \begin{center}
    \includegraphics[width=0.6\textwidth]{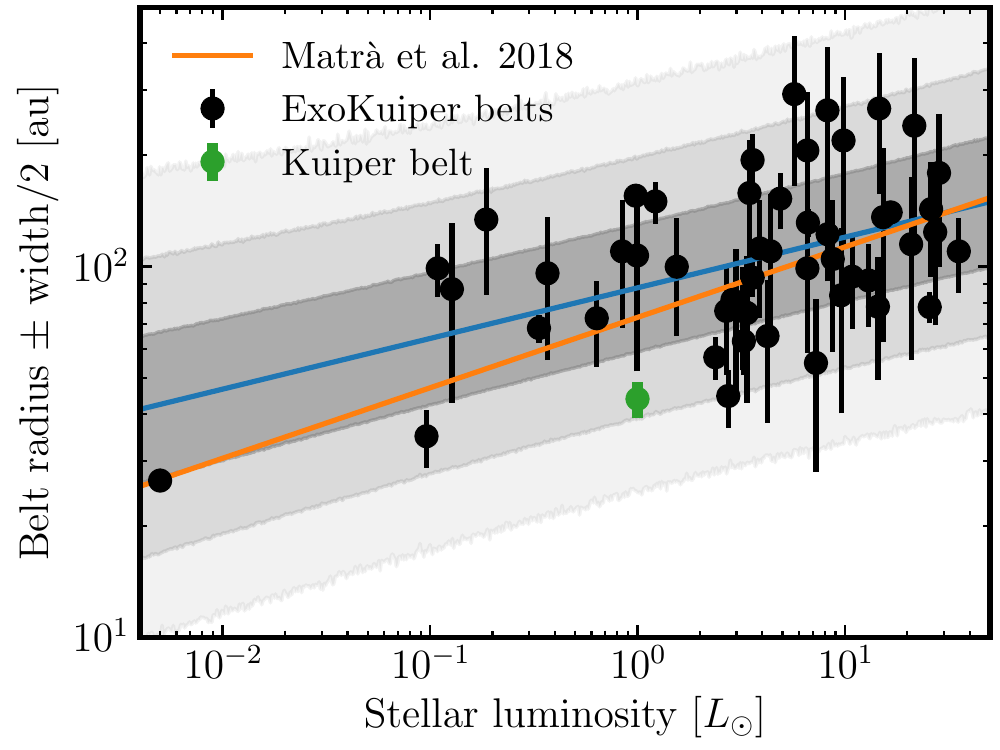}
    \caption{ExoKuiper belts radii (black) vs stellar luminosity from
      the REASONS survey and \citet{Cronin-Coltsmann2021}. The Kuiper
      belt is shown in green. The errorbars represent half of the
      full-width-half-maximum. The orange line shows the previous best
      fit in \citet{Matra2018mmlaw}, while the blue line shows the new
      best fit. The grey shaded regions represent the 68\%, 95\% and
      99.7\% confidence regions taking into account a fit to the
      intrinsic dispersion (assumed to be constant across stellar
      luminosity).} \label{fig:mmlaw}
  \end{center}
\end{figure}

The radius luminosity relation motivated the REsolved ALMA and SMA
Observations of Nearby Stars (REASONS) survey, which is the follow up
of the JCMT SONS survey \citep{Holland2017}. The aim of REASONS was to
double the sample size of millimeter-resolved debris discs, and model
all discs (including archival observations) to derive their radius and
width in a consistent manner. Preliminary results shown in Figure
\ref{fig:mmlaw} \citep[][Matr\`a et al. in prep.]{Sepulveda2019},
including one additional observation of Fomalhaut C
\citep{Cronin-Coltsmann2021}, indicate that the trend still holds in a
sample of 52 systems.

The new preliminary results show a significantly larger dispersion,
but the trend is still significant (blue line). Note that similar
trends are known in protoplanetary discs, with discs around more
massive stars being larger on average \citep{Andrews2018sizes}. It is
still an open question if these two relations are linked or if the
location of exoKuiper belts is connected with the CO snow line
location. Finally, it is worth noting as well that the Kuiper belt is
small compared to the majority of belts around FGK stars, although the
exoKuiper belts in this sample are much more massive than the Kuiper
belt due to the sensitivity limit of current instruments. Smaller
exoKuiper belts could be intrinsically less massive and fainter.

\subsubsection{Widths}
\label{sec:widths}

\begin{figure}[h!]
  \begin{center}
    \includegraphics[width=0.6\textwidth]{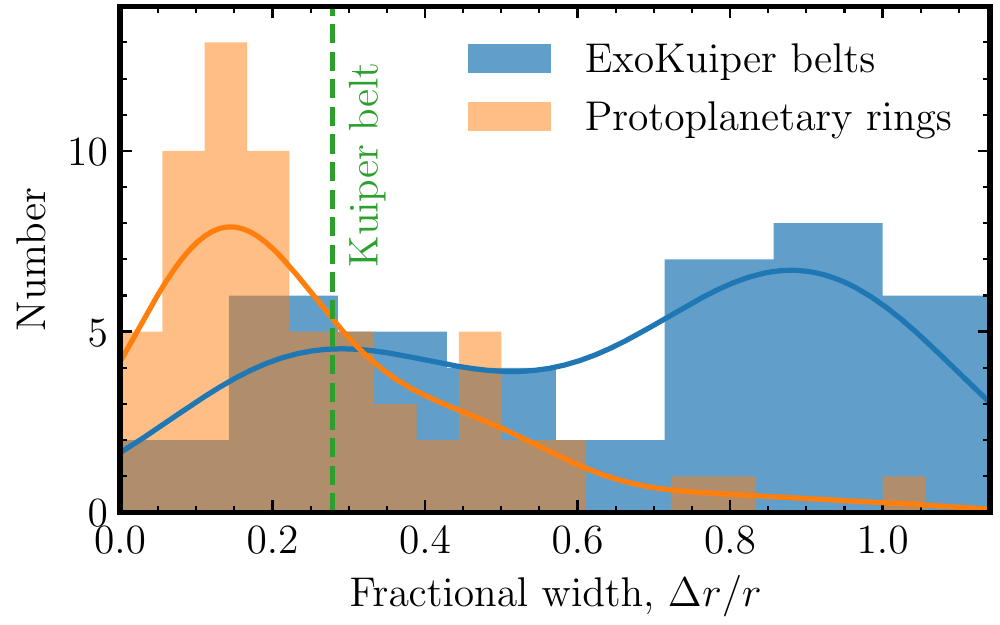}
    \caption{Histogram of the fractional width of exoKuiper belts from
      the REASONS survey (blue, Matr\'a et al. in prep) and rings in
      protoplanetary discs \citep[orange,][]{Huang2018, Long2018,
        Cieza2021}. The green dashed line represents the fractional
      width of the Kuiper belt estimated using the
      full-width-half-maximum of the L7 synthetic model of the inner,
      main and outer Kuiper belt \citep{Kavelaars2009, Petit2011}. The
      solid lines represent kernel density estimations using a
      Gaussian kernel. } \label{fig:dr}
  \end{center}
\end{figure}

Together with the radius information, REASONS has also constrained the
width of $\sim$40 belts. The results show that most belts are
relatively wide, with a median fractional width (the ratio between the
width and radius) of 0.74. Figure~\ref{fig:dr} shows the distribution
of fractional widths of exoKuiper belts (blue), revealing that there
is not a typical width, but rather a wide distribution. Narrow belts
like the one around Fomalhaut are rare. In fact, even the Kuiper belt
seems to be much narrower than most belts.

Figure~\ref{fig:dr} also shows the fractional widths of dust rings in
protoplanetary discs with radii between 30 to 250 au (orange). Those
rings occupy the same range of radii as exoKuiper belts, but are much
narrower with a median fractional width of 0.18. This difference is
interesting, since those dusty rings are ideal places for planetesimal
formation \citep{Stammler2019, Carrera2021}, and thus we would expect
exoKuiper belts to be similar or even narrower. Nevertheless, as
recently shown by \citet{Miller2021}, the large width of exoKuiper
belts could be explained if those rings migrate forming planetesimals
at a wide range of radii. Those rings would naturally migrate if, for
example, they were caused by planets. In that case, the width of
debris discs would imply migration rates higher than 5~au~Myr$^{-1}$
in order to cover distances of 50-100~au in the 5-10 Myr
protoplanetary disc lifetimes.

\subsubsection{Radial structures}

\begin{figure}[h!]
  \begin{center}
    \includegraphics[width=1.0\textwidth]{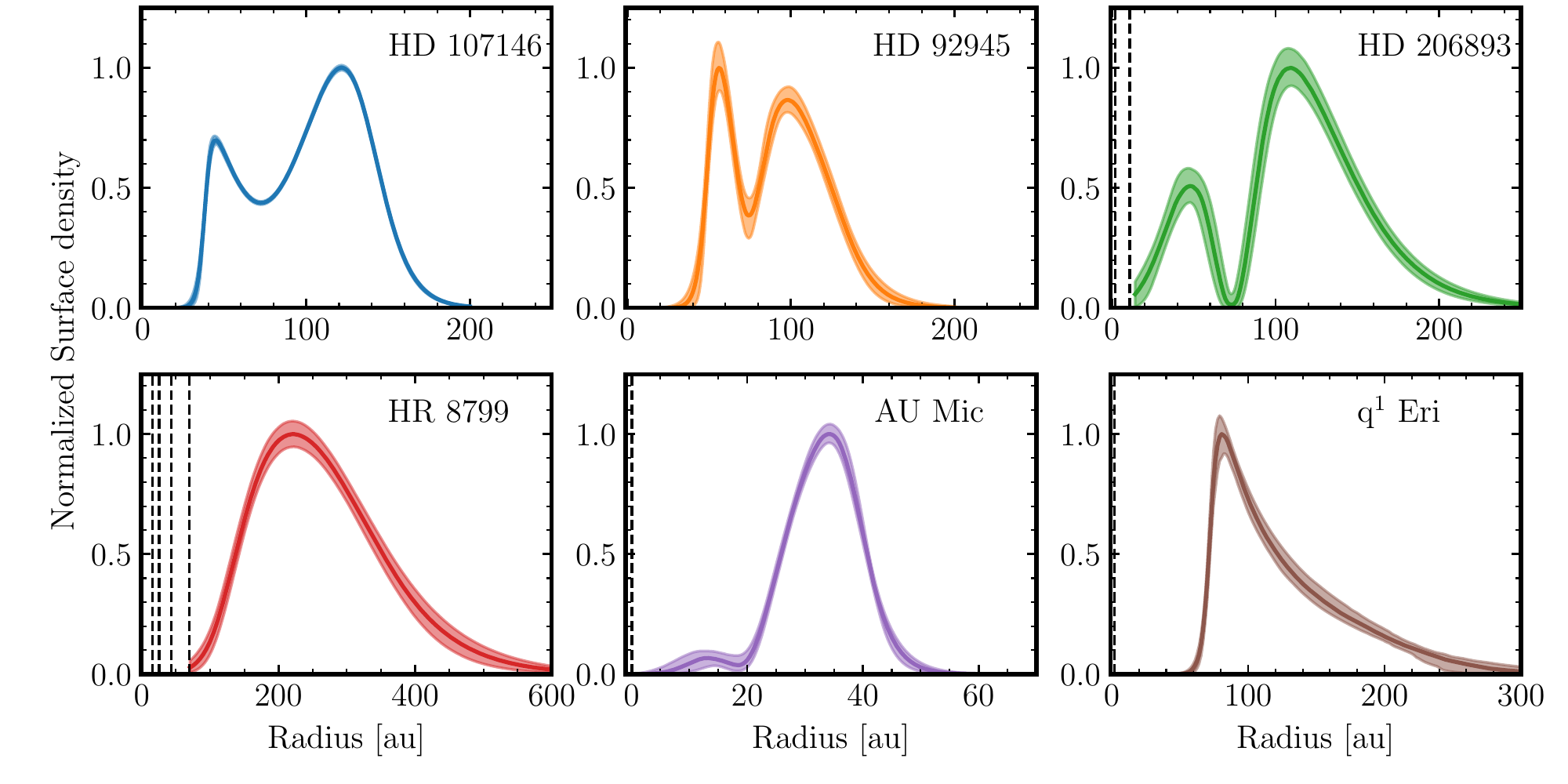}
    \caption{Surface density radial profiles of 6 exoKuiper belts
      observed with ALMA. The profiles were retrieved using parametric
      models that were sensitive to the presence of gaps and the
      smoothness of the inner and outer edges
      \citep[][]{Marino2021}. The shaded regions correspond to the
      68\% confidence interval. The vertical dashed lines show the
      location of known planets in these systems.} \label{fig:sr}
  \end{center}
\end{figure}

In addition to measuring the central radius and widths, ALMA's
resolving power has uncovered the radial substructures in multiple
wide exoKuiper belts. Figure~\ref{fig:sr} shows the surface density
profiles of six belts, which display different features. For example,
HD~107146, HD~92945, HD~206893 and possibly AU~Mic show gaps
\citep{Marino2018hd107, Marino2019, Marino2020hd206, Nederlander2021,
  Daley2019} that could have been carved by planets orbiting within
those gaps through scattering \citep[]{Morrison2015}. If those planets
have not migrated, the width of those gaps ($\Delta_{\rm gap}$) should
be roughly equal to the chaotic zone \citep{Wisdom1980},
i.e. $\Delta_{\rm gap}\approx 3 a_{plt} (M_{\rm
  plt}/M_{\star})^{2/7}$, where $a_{plt}$, $M_{\rm plt}$ and
$M_{\star}$ are the planet semi-major axis, its mass, and the stellar
mass, respectively. The gaps in HD~92945 and HD~206893 are centred at
roughly 75~au and span $\sim15-30$~au, which would suggest planet
masses in between Neptune's and Jupiter's. HD~107146's gap is very
wide, but not deep enough to be explained by a single planet. It is
also possible that the gaps were open by planets closer in through
secular interactions \citep{Pearce2015doublering, Yelverton2018,
  Sefilian2020}, in which case they would likely be asymmetric (which
has only been excluded for HD~107146). Gaps have been also been found
in scattered light observations tracing smaller grains
\citep{Golimowski2011, Schneider2014, Engler2019, Perrot2016,
  Bonnefoy2017, Feldt2017, Boccaletti2019, Ren2021}. Gaps in the
distribution of small grains could be generated by the photoelectric
instability if gas is present \citep{Lyra2013photoelectric,
  Richert2017}, which is the case in some of these discs
(e.g. HD~141569 and HD~131835, see \S\ref{sec:gas}).

Another interesting feature is the inner and outer edges of the
belts. As discussed in \S\ref{sec:coll}, sharp inner edges might
indicate that discs were truncated by an interior planet. On the other
hand, outer edges can be used to constrain the level of dynamical
excitation of a disc as the higher the dispersion of eccentricities
is, the smoother the edge will be \citep{Marino2021}. Among the 6
discs shown in Figure~\ref{fig:sr}, HD~206803, HR~8799 and q$^{1}$~Eri
have the smoothest outer edges. The first two are known to host gas
giant planets just interior to their discs \citep{Marois2010,
  Milli2017}, which could have scattered a large amount of solids onto
eccentric orbits producing smooth outer edges, similar to the
scattered component of the Kuiper belt. Scattered populations could
explain as well the halos seen in ALMA observations of a few discs
\citep{Marino2016, MacGregor2018}.

Finally, there are also some narrow belts that are known to be
eccentric \citep[e.g. Fomalhaut, HR~4796, HD~202628, HD~15115,
  HD~106906,][]{Kalas2005, Rodigas2015, Krist2012, Sai2015,
  Kalas2015}. This has normally been attributed to the presence of an
eccentric planet inducing an eccentricity in the belt through secular
interactions \citep{Wyatt1999, Quillen2006, Chiang2009, Nesvold2013,
  Pearce2014}. However, \citet{Kennedy2020} recently showed that two
of these discs (Fomalhaut and HD~202628) appear narrower than expected
in that scenario, suggesting that either the belts were born eccentric
or eccentricities are damped leading to narrower belts.

%% such as
%% Fomalhaut, are known to be eccentric
%% \subsection{Other structures}
%% eccentric discs and Kennedy+2020, asymmetric gaps, etc

\subsubsection{Vertical structures}

The vertical structure of exoKuiper belts can also provide key
information about the dynamics of these systems. Studies of near to
edge-on belts are ideally suited for this. The best example is perhaps
$\beta$~Pic, which has a large and bright edge-on disc. HST and
ground-based scattered light images of this disc showed that it is is
warped \citep{Mouillet1997, Heap2000, Golimowski2006}, which was
attributed to an inner companion misaligned to the disc midplane
\citep{Mouillet1997, Augereau2001}. This putative planet was later
discovered through direct imaging \citep{Lagrange2009}, in rough
agreement with the misaligned scenario \citep{Dawson2011,
  Nesvold2015betapic}.

Furthermore, the vertical distribution of the disc is significantly
deviated from a simple Gaussian distribution \citep{Golimowski2006,
  Matra2019betapic}. In a disc of interacting planetesimals, the
distribution of inclinations is expected to follow a Rayleigh
distribution \citep{Ida1992}, which translates to a Gaussian particle
density distribution. However, $\beta$~Pic's disc is better described
by two Gaussians. \citet{Matra2019betapic} argued that the population
with high inclinations could result as a consequence of an additional
planet migrating out. This scenario would be similar to Neptune's
outward migration that led to the two dynamical populations in the
classical Kuiper belts \citep{Brown2001, Nesvorny2015}.

AU~Mic is another well known system with an edge-on disc around an M
star. Using ALMA, \citet{Daley2019} showed that the disc height is
consistent with an inclination dispersion of $\sim3^{\circ}$ and that
such a excitation could be produced by planetesimals larger than
400~km embedded in the disc. This disc also shows fast outward moving
clumps in scattered light, some of which seem to be unbound
\citep{Boccaletti2018}. These clumps would be made of small grains
that are on unbound trajectories due to strong stellar winds by this
young M star. However, what creates the clumps is still highly
uncertain. Proposed scenarios range from a compact body embedded in
the belt releasing dust \citep{Sezestre2017}, a collisional avalanche
at the intersection of two belts \citep{Chiang2017}, or asymmetric
disc winds \citep{Wisniewski2019}.

\section{Circumstellar gas}
\label{sec:gas}
While debris discs are generally considered gas poor, it has been
known for more than two decades that some debris discs systems have
gas that is readily detectable as absorption lines in the UV and
optical, or as emission lines at far-IR and mm wavelengths
\citep{Slettebak1975, Zuckerman1995}. Historically, these two type of
observations have been studied as independent phenomena, although
several systems present both. Below, I summarise the main
characteristics of these lines and their potential origins and
implications.

\subsection{Absorption lines}

The first known and best studied system with circumstellar absorption
lines is $\beta$~Pic. \citet{Slettebak1975} found narrow CaII H\&K
absorption lines in the stellar spectrum. Subsequent studies found
that these and FeII lines in the UV change with time \citep{Kondo1985,
  Lagrange1987}. \citet{Ferlet1987} showed how these absorption lines
can be decomposed into stable narrow lines at the stellar velocity
and Doppler shifted lines that vary in timescales of months, days and
hours. These variations were attributed to the presence of star-grazer
objects or \textit{falling evaporating bodies} on highly eccentric
orbits that sublimate within a few stellar radii producing the
observed warm gas \citep{Beust1990, Kiefer2014b}. More recently,
\citet{Kennedy2018} showed how individual absorption features
accelerate over several hours, and this acceleration constrains the
orbits of these \textit{exocomets}. Today we know that this phenomenon
is common, and more likely to be detected in systems with debris discs
that are edge-on \citep[e.g.][]{Welsh1998, Eiroa2016, Hales2017,
  Iglesias2018, Rebollido2020}. For a more in depth description of the
gas seen in absorption and its comparison with Solar System comets see
the recent review by \citet{Strom2020}.

%% Around more massive stars, these absorption lines  even idicate

\subsection{Emission lines}

Colder gas seen in emission at far-IR and mm wavelengths have also been
found in debris discs. At first, this gas was mainly found around
young A stars with bright debris discs \citep[e.g.][]{Zuckerman1995,
  Kospal2013, Cataldi2014, Dent2014, Moor2017}, but deeper
observations, especially with ALMA, have shown that CO gas can also
been found in older systems \citep{Matra2017fomalhaut}, and later type
stars \citep{Marino2016, Marino2017etacorvi, Matra2019twa7, Kral2020COsurvey}.

The origin of this gas in many of these systems is still a subject of
debate. \citet{Zuckerman2012} hypothesised that the CO gas could be
released by the same planetesimals that resupply the dust if these
were volatile rich (i.e. exocomets). As solids fragment due to
collisions, CO gas that was trapped in pockets could be release, and
thus debris discs should release exocometary gas as they collisionally
evolve. The rate at which gas is released would be a fraction of the
disc mass loss rate, and this fraction should be close to the
abundance of volatiles in exocomets. However, \citet{Zuckerman2012}
realised the gas release rate would not be enough to balance the rate
at which CO was being destroyed. If unshielded, CO molecules quickly
photodissociate due to stellar or interstellar UV photons in
timescales of $\lesssim100$~yr \citep{Visser2009}. This meant that
this \textit{secondary origin} scenario could explain the low CO
levels in some systems and provide constraints on the abundance of CO
in their exocomets \citep[e.g. HD~181327, $\beta$~Pic and Fomalhaut it
  was found to be consistent with Solar System comets,][]{Marino2016,
  Kral2016, Matra2017fomalhaut}, but not the high CO levels around
some young A stars. It was concluded that the CO gas in those stars
was likely a leftover from their protoplanetary discs and shielded by
leftover H$_2$ gas. These systems were tagged as hybrid, containing
secondary dust and primordial gas \citep{Kospal2013}.

This hybrid nature was recently challenged by \citet{Kral2019} that
proposed that the same carbon produced by the CO photodissociation
could shield CO from the interstellar UV, increasing the CO lifetime
by orders of magnitude. This longer lifetime due to carbon shielding
would explain the high CO levels around young A stars and would only
happen to those that are born with the most massive debris discs
\citep[][]{Marino2020gas}, disfavouring FGK stars that tend to have
lower mass discs. More recent observations have found abundant carbon
gas in some of the systems with high CO levels that are expected to be
shielded \citep{Kral2019, Higuchi2019, Cataldi2020}, although is is
still an open question if the observed carbon levels are high
enough. The gas evolution and shielding of CO gas are described below.

\subsection{Evolution of exocometary gas}

Once this gas is released from planetesimals it is expected to spread
forming an accretion disc \citep{Kral2016}, unless radiation pressure
or stellar winds are strong enough to blow out the gas
\citep{Youngblood2021, Kral2021}. How massive such a gaseous
circumstellar disc can be depends on the rate at which gas is being
released and the rate at which gas is lost via accretion onto the
central star or inner bodies. The accretion rate will depend on the
efficiency of the angular momentum transport, either through a
kinematic viscosity (e.g. if the magnetorotational instability is
active) or disc winds \citep{Kral2016b}. So far, models assume the
disc is evolving viscously and parametrise the viscosity through the
classical alpha disc model with $\nu=\alpha c_s^2/\Omega_{\rm K}$,
where $c_s$, $\Omega_{\rm K}$ and $\alpha$ are the sound speed,
Keplerian frequency and an dimensionless free parameter smaller than
one. A population synthesis study by \citet{Marino2020gas} found a
good agreement between models and observations with $\alpha\sim0.1$,
although its value still is very uncertain given the assumptions in
these models. Below I describe how the gas is expected to viscously
evolve.

\begin{figure}[b]
  \begin{center}
    \includegraphics[width=1.0\textwidth]{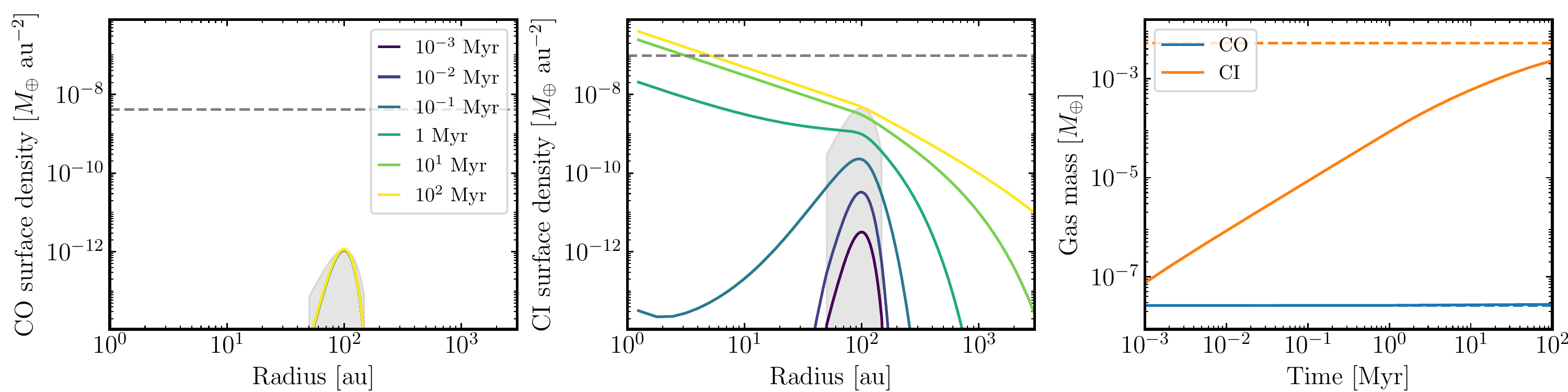}
    \includegraphics[width=1.0\textwidth]{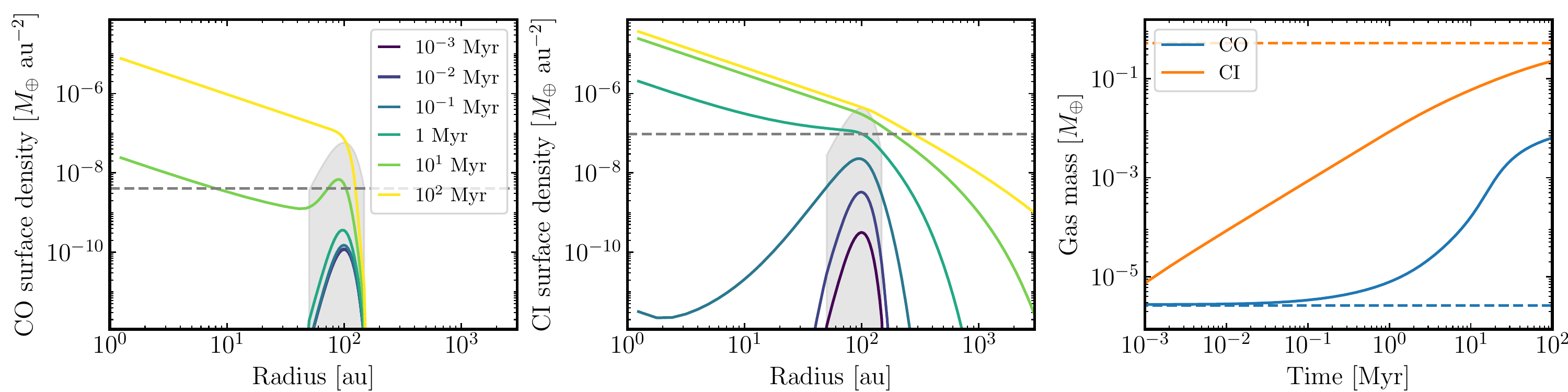}
    \caption{Evolution of circumstellar CO and CI gas as it is
      released from a debris disc at 100~au according to the python
      package
      \href{https://github.com/SebaMarino/exogas}{\textsc{exogas}}
      \citep{Marino2020gas}. The left and middle panels show the
      surface density of CO and CI at different epochs, respectively.
      The right panels show the evolution of the total CO and CI
      mass. The grey dashed lines represent the surface density at
      which the CO and CI become optically thick to photodissociating
      UV photons in the vertical direction. The blue and orange dashed
      lines show the expected mass in steady-state if CO is
      unshielded.} \label{fig:gas_evol}
  \end{center}
\end{figure}

Once gas is released in the form of CO, it will photodissociate into
carbon and oxygen atoms. Since the CO lifetime is only about
$\sim100$~yr, it will stay co-located with the dust and
planetesimals. Carbon and oxygen, on the other hand, will accumulate
and viscously spread on longer timescales. This behaviour is shown in
the top panels of Figure~\ref{fig:gas_evol}, which assume CO gas is
released at a rate of $2\times10^{-4}\ M_{\oplus}$~Myr$^{-1}$ from a
belt of planetesimals centred at 100~au and 50~au wide around a 10
$L_{\odot}$ star, $\alpha=10^{-2}$, and carbon is mostly neutral (only
neutral carbon can shield CO). With these values, neutral carbon (CI)
never reaches the critical surface density of
$10^{-7}\ M_{\oplus}$~au$^{-2}$ (grey dashed line) and thus it is
ineffective at shielding CO. This is why the CO surface density stays
roughly constant over 100~Myr and does not spread beyond the belt. If
the gas release rate is higher,
e.g. $2\times10^{-2}\ M_{\oplus}$~Myr$^{-1}$ (as shown in bottom
panels of Figure~\ref{fig:gas_evol}), CI reaches the critical surface
density, after which CO becomes shielded and viscously spreads to the
inner and outer regions. These simulations were done with the python
package
\href{https://github.com/SebaMarino/exogas}{\textsc{exogas}}\footnote{\textsc{exogas}
can be downloaded from https://github.com/SebaMarino/exogas. If you
would like to use it in a publication, please get in touch.}  that is
based on the software developed in \citet{Marino2020gas}, with the
addition of photon counting to correctly calculate the
photodissociation timescale of CO \citep{Cataldi2020}.

Note that these simulations assume that most of carbon sits in an
upper layer on top of the CO gas, however, if they are well mixed
shielding by carbon becomes much less effective
\citep{Cataldi2020}. Therefore, observations that can constrain the
vertical distribution of CO and carbon gas are required to assess how
effective carbon is at shielding CO. Finally, although these
simulations assume a constant CO release rate, this is expected to
decrease with time, which means the gas levels can be out of
equilibrium and quickly evolving from a shielded onto an unshielded
state \citep{Marino2020gas}.

\subsection{Implications}

Understanding the evolution of this exocometary gas is important for
several reasons. First, the chemical composition of this gas encodes
the composition of planetesimals. Therefore, by understanding how this
gas evolves it would be possible to constrain the volatile composition
of exocomets, even in shielded discs. Second, as it viscously spreads,
the gas could be accreted by gas giant planets in the inner regions
and potentially change their abundances of volatile elements in their
atmosphere \citep{Marino2020gas}. This would make even more difficult
to trace the formation location of gas giants based on their
atmospheric abundances \citep{Oberg2011}. Finally, the exocometary gas
could also be accreted by low mass planets in the habitable zone,
building new secondary atmospheres with exocometary composition over
tens of Myr's \citep{Kral2020atmospheres}.

\noindent {\bf Acknowledgements:} I would like to thank the organisers
of the Severo-Ochoa advanced school ``Planets, exoplanets and their
systems in a broad and multidisciplinary context'', which this review
chapter is based on. I would also like to thank Grant Kennedy, Luca
Matr\`a and Josh Lovell for providing ALMA images of Fomalhaut,
$\beta$~Pic and q$^{1}$ Eri.

\bibliographystyle{shortbib_v2} % this creates shorter references
\bibliography{SM_pformation.bib} % if your bibtex file is called example.bib

\begin{thebibliography}{201}
\providecommand{\natexlab}[1]{#1}

\bibitem[\protect\astroncite{{Absil} et~al.}{2021}]{Absil2021}
{Absil}, O; et~al., 2021.
\newblock \aap, 651:A45.

\bibitem[\protect\astroncite{{Ahmic} et~al.}{2009}]{Ahmic2009}
{Ahmic}, M; et~al., 2009.
\newblock \apj, 705:529.

\bibitem[\protect\astroncite{{Andrews} et~al.}{2018}]{Andrews2018sizes}
{Andrews}, SM; et~al., 2018.
\newblock \apj, 865:157.

\bibitem[\protect\astroncite{{Augereau} et~al.}{2001}]{Augereau2001}
{Augereau}, JC; et~al., 2001.
\newblock \aap, 370:447.

\bibitem[\protect\astroncite{{Aumann} et~al.}{1984}]{Aumann1984}
{Aumann}, HH; et~al., 1984.
\newblock \apjl, 278:L23.

\bibitem[\protect\astroncite{{Backman} et~al.}{2009}]{Backman2009}
{Backman}, D; et~al., 2009.
\newblock \apj, 690:1522.

\bibitem[\protect\astroncite{{Backman} \& {Paresce}}{1993}]{Backman1993}
{Backman}, DE et~al., 1993.
\newblock In EH~{Levy} \& JI~{Lunine}, eds., \emph{Protostars and Planets III}.
  1253--1304.

\bibitem[\protect\astroncite{{Ballering} et~al.}{2014}]{Ballering2014}
{Ballering}, NP; et~al., 2014.
\newblock \apj, 793:57.

\bibitem[\protect\astroncite{{Ballering} et~al.}{2016}]{Ballering2016}
{Ballering}, NP; et~al., 2016.
\newblock \apj, 823:108.

\bibitem[\protect\astroncite{{Balog} et~al.}{2009}]{Balog2009}
{Balog}, Z; et~al., 2009.
\newblock \apj, 698:1989.

\bibitem[\protect\astroncite{{Beust} et~al.}{1990}]{Beust1990}
{Beust}, H; et~al., 1990.
\newblock \aap, 236:202.

\bibitem[\protect\astroncite{Blum \& Wurm}{2008}]{Blum2008}
Blum, J et~al., 2008.
\newblock Annual Review of Astronomy and Astrophysics, 46:21.

\bibitem[\protect\astroncite{{Boccaletti} et~al.}{2018}]{Boccaletti2018}
{Boccaletti}, A; et~al., 2018.
\newblock \aap, 614:A52.

\bibitem[\protect\astroncite{{Boccaletti} et~al.}{2019}]{Boccaletti2019}
{Boccaletti}, A; et~al., 2019.
\newblock \aap, 625:A21.

\bibitem[\protect\astroncite{{Bonnefoy} et~al.}{2017}]{Bonnefoy2017}
{Bonnefoy}, M; et~al., 2017.
\newblock \aap, 597:L7.

\bibitem[\protect\astroncite{{Bonsor} et~al.}{2012}]{Bonsor2012nbody}
{Bonsor}, A; et~al., 2012.
\newblock \aap, 548:A104.

\bibitem[\protect\astroncite{{Bonsor} \& {Wyatt}}{2012}]{Bonsor2012analytic}
{Bonsor}, A et~al., 2012.
\newblock \mnras, 420:2990.

\bibitem[\protect\astroncite{{Bonsor} et~al.}{2013}]{Bonsor2013}
{Bonsor}, A; et~al., 2013.
\newblock \mnras, 433:2938.

\bibitem[\protect\astroncite{{Bonsor} et~al.}{2014}]{Bonsor2014}
{Bonsor}, A; et~al., 2014.
\newblock \mnras, 441:2380.

\bibitem[\protect\astroncite{{Bonsor} et~al.}{2018}]{Bonsor2018}
{Bonsor}, A; et~al., 2018.
\newblock \mnras, 480:5560.

\bibitem[\protect\astroncite{{Booth} et~al.}{2013}]{Booth2013}
{Booth}, M; et~al., 2013.
\newblock \mnras, 428:1263.

\bibitem[\protect\astroncite{{Brown}}{2001}]{Brown2001}
{Brown}, ME, 2001.
\newblock \aj, 121:2804.

\bibitem[\protect\astroncite{{Carrera} et~al.}{2021}]{Carrera2021}
{Carrera}, D; et~al., 2021.
\newblock \aj, 161:96.

\bibitem[\protect\astroncite{{Casassus} et~al.}{2019}]{Casassus2019}
{Casassus}, S; et~al., 2019.
\newblock \mnras, 483:3278.

\bibitem[\protect\astroncite{{Cataldi} et~al.}{2014}]{Cataldi2014}
{Cataldi}, G; et~al., 2014.
\newblock \aap, 563:A66.

\bibitem[\protect\astroncite{{Cataldi} et~al.}{2020}]{Cataldi2020}
{Cataldi}, G; et~al., 2020.
\newblock \apj, 892:99.

\bibitem[\protect\astroncite{{Chambers}}{2013}]{Chambers2013}
{Chambers}, JE, 2013.
\newblock \icarus, 224:43.

\bibitem[\protect\astroncite{{Chen} et~al.}{2009}]{Chen2009}
{Chen}, CH; et~al., 2009.
\newblock \apj, 701:1367.

\bibitem[\protect\astroncite{{Chiang} et~al.}{2009}]{Chiang2009}
{Chiang}, E; et~al., 2009.
\newblock \apj, 693:734.

\bibitem[\protect\astroncite{{Chiang} \& {Fung}}{2017}]{Chiang2017}
{Chiang}, E et~al., 2017.
\newblock \apj, 848:4.

\bibitem[\protect\astroncite{{Choquet} et~al.}{2018}]{Choquet2018}
{Choquet}, {\'E}; et~al., 2018.
\newblock \apj, 854:53.

\bibitem[\protect\astroncite{{Cieza} et~al.}{2021}]{Cieza2021}
{Cieza}, LA; et~al., 2021.
\newblock \mnras, 501:2934.

\bibitem[\protect\astroncite{{Cronin-Coltsmann}
  et~al.}{2021}]{Cronin-Coltsmann2021}
{Cronin-Coltsmann}, PF; et~al., 2021.
\newblock \mnras, 504:4497.

\bibitem[\protect\astroncite{{Daley} et~al.}{2019}]{Daley2019}
{Daley}, C; et~al., 2019.
\newblock \apj, 875:87.

\bibitem[\protect\astroncite{{Dawson} et~al.}{2011}]{Dawson2011}
{Dawson}, RI; et~al., 2011.
\newblock \apjl, 743:L17.

\bibitem[\protect\astroncite{{Dent} et~al.}{2014}]{Dent2014}
{Dent}, WRF; et~al., 2014.
\newblock Science, 343:1490.

\bibitem[\protect\astroncite{{Dohnanyi}}{1969}]{Dohnanyi1969}
{Dohnanyi}, JS, 1969.
\newblock \jgr, 74:2531.

\bibitem[\protect\astroncite{{Dominik} \& {Decin}}{2003}]{Dominik2003}
{Dominik}, C et~al., 2003.
\newblock \apj, 598:626.

\bibitem[\protect\astroncite{{Dullemond} et~al.}{2018}]{Dullemond2018dsharp}
{Dullemond}, CP; et~al., 2018.
\newblock \apjl, 869:L46.

\bibitem[\protect\astroncite{{Eiroa} et~al.}{2013}]{Eiroa2013}
{Eiroa}, C; et~al., 2013.
\newblock \aap, 555:A11.

\bibitem[\protect\astroncite{{Eiroa} et~al.}{2016}]{Eiroa2016}
{Eiroa}, C; et~al., 2016.
\newblock \aap, 594:L1.

\bibitem[\protect\astroncite{{Engler} et~al.}{2019}]{Engler2019}
{Engler}, N; et~al., 2019.
\newblock \aap, 622:A192.

\bibitem[\protect\astroncite{{Ertel} et~al.}{2020}]{Ertel2020}
{Ertel}, S; et~al., 2020.
\newblock \aj, 159:177.

\bibitem[\protect\astroncite{{Esposito} et~al.}{2020}]{Esposito2020}
{Esposito}, TM; et~al., 2020.
\newblock \aj, 160:24.

\bibitem[\protect\astroncite{{Faramaz} et~al.}{2017}]{Faramaz2017}
{Faramaz}, V; et~al., 2017.
\newblock \mnras, 465:2352.

\bibitem[\protect\astroncite{{Faramaz} et~al.}{2021}]{Faramaz2021}
{Faramaz}, V; et~al., 2021.
\newblock \aj, 161:271.

\bibitem[\protect\astroncite{{Feldt} et~al.}{2017}]{Feldt2017}
{Feldt}, M; et~al., 2017.
\newblock \aap, 601:A7.

\bibitem[\protect\astroncite{{Ferlet} et~al.}{1987}]{Ferlet1987}
{Ferlet}, R; et~al., 1987.
\newblock \aap, 185:267.

\bibitem[\protect\astroncite{{G{\'a}sp{\'a}r} et~al.}{2012}]{Gaspar2012}
{G{\'a}sp{\'a}r}, A; et~al., 2012.
\newblock \apj, 754:74.

\bibitem[\protect\astroncite{{Genda} et~al.}{2015}]{Genda2015}
{Genda}, H; et~al., 2015.
\newblock \apj, 810:136.

\bibitem[\protect\astroncite{{Golimowski} et~al.}{2006}]{Golimowski2006}
{Golimowski}, DA; et~al., 2006.
\newblock \aj, 131:3109.

\bibitem[\protect\astroncite{{Golimowski} et~al.}{2011}]{Golimowski2011}
{Golimowski}, DA; et~al., 2011.
\newblock \aj, 142:30.

\bibitem[\protect\astroncite{{Graham} et~al.}{2007}]{Graham2007}
{Graham}, JR; et~al., 2007.
\newblock \apj, 654:595.

\bibitem[\protect\astroncite{{Hales} et~al.}{2017}]{Hales2017}
{Hales}, AS; et~al., 2017.
\newblock \mnras, 466:3582.

\bibitem[\protect\astroncite{{Heap} et~al.}{2000}]{Heap2000}
{Heap}, SR; et~al., 2000.
\newblock \apj, 539:435.

\bibitem[\protect\astroncite{{Higuchi} et~al.}{2019}]{Higuchi2019}
{Higuchi}, AE; et~al., 2019.
\newblock \apj, 883:180.

\bibitem[\protect\astroncite{{Holland} et~al.}{2017}]{Holland2017}
{Holland}, WS; et~al., 2017.
\newblock \mnras, 470:3606.

\bibitem[\protect\astroncite{{Huang} et~al.}{2018}]{Huang2018}
{Huang}, J; et~al., 2018.
\newblock \apjl, 869:L42.

\bibitem[\protect\astroncite{{Hughes} et~al.}{2018}]{Hughes2018}
{Hughes}, AM; et~al., 2018.
\newblock \araa, 56:541.

\bibitem[\protect\astroncite{{Ida} \& {Makino}}{1992}]{Ida1992}
{Ida}, S et~al., 1992.
\newblock \icarus, 96:107.

\bibitem[\protect\astroncite{{Iglesias} et~al.}{2018}]{Iglesias2018}
{Iglesias}, D; et~al., 2018.
\newblock \mnras, 480:488.

\bibitem[\protect\astroncite{{Jackson} \& {Wyatt}}{2012}]{Jackson2012}
{Jackson}, AP et~al., 2012.
\newblock \mnras, 425:657.

\bibitem[\protect\astroncite{{Janson} et~al.}{2021}]{Janson2021}
{Janson}, M; et~al., 2021.
\newblock \aap, 646:A132.

\bibitem[\protect\astroncite{{Johansen} et~al.}{2007}]{Johansen2007}
{Johansen}, A; et~al., 2007.
\newblock \nat, 448:1022.

\bibitem[\protect\astroncite{{Kalas} et~al.}{2005}]{Kalas2005}
{Kalas}, P; et~al., 2005.
\newblock \nat, 435:1067.

\bibitem[\protect\astroncite{{Kalas} et~al.}{2015}]{Kalas2015}
{Kalas}, PG; et~al., 2015.
\newblock \apj, 814:32.

\bibitem[\protect\astroncite{{Kataoka} et~al.}{2013}]{Kataoka2013}
{Kataoka}, A; et~al., 2013.
\newblock \aap, 557:L4.

\bibitem[\protect\astroncite{{Kavelaars} et~al.}{2009}]{Kavelaars2009}
{Kavelaars}, JJ; et~al., 2009.
\newblock \aj, 137:4917.

\bibitem[\protect\astroncite{{Kennedy}}{2020}]{Kennedy2020}
{Kennedy}, GM, 2020.
\newblock Royal Society Open Science, 7:200063.

\bibitem[\protect\astroncite{{Kennedy} \& {Wyatt}}{2010}]{Kennedy2010}
{Kennedy}, GM et~al., 2010.
\newblock \mnras, 405:1253.

\bibitem[\protect\astroncite{{Kennedy} \& {Wyatt}}{2014}]{Kennedy2014}
{Kennedy}, GM et~al., 2014.
\newblock \mnras, 444:3164.

\bibitem[\protect\astroncite{{Kennedy} et~al.}{2015}]{Kennedy2015lbti}
{Kennedy}, GM; et~al., 2015.
\newblock \apjs, 216:23.

\bibitem[\protect\astroncite{{Kennedy} et~al.}{2018}]{Kennedy2018}
{Kennedy}, GM; et~al., 2018.
\newblock \mnras, 475:4924.

\bibitem[\protect\astroncite{{Kenyon} \& {Bromley}}{2008}]{Kenyon2008}
{Kenyon}, SJ et~al., 2008.
\newblock \apjs, 179:451-483.

\bibitem[\protect\astroncite{{Kiefer} et~al.}{2014}]{Kiefer2014b}
{Kiefer}, F; et~al., 2014.
\newblock \nat, 514:462.

\bibitem[\protect\astroncite{{Kirchschlager} et~al.}{2017}]{Kirchschlager2017}
{Kirchschlager}, F; et~al., 2017.
\newblock \mnras, 467:1614.

\bibitem[\protect\astroncite{{K{\"o}hler} et~al.}{2008}]{Kohler2008}
{K{\"o}hler}, M; et~al., 2008.
\newblock \apjl, 686:L95.

\bibitem[\protect\astroncite{{Kondo} \& {Bruhweiler}}{1985}]{Kondo1985}
{Kondo}, Y et~al., 1985.
\newblock \apjl, 291:L1.

\bibitem[\protect\astroncite{{K{\'o}sp{\'a}l} et~al.}{2013}]{Kospal2013}
{K{\'o}sp{\'a}l}, {\'A}; et~al., 2013.
\newblock \apj, 776:77.

\bibitem[\protect\astroncite{{Kral} et~al.}{2016}]{Kral2016}
{Kral}, Q; et~al., 2016.
\newblock \mnras, 461:845.

\bibitem[\protect\astroncite{{Kral} \& {Latter}}{2016}]{Kral2016b}
{Kral}, Q et~al., 2016.
\newblock \mnras, 461:1614.

\bibitem[\protect\astroncite{{Kral} et~al.}{2017}]{Kral2017exozodis}
{Kral}, Q; et~al., 2017.
\newblock The Astronomical Review, 13:69.

\bibitem[\protect\astroncite{{Kral} et~al.}{2019}]{Kral2019}
{Kral}, Q; et~al., 2019.
\newblock \mnras, 489:3670.

\bibitem[\protect\astroncite{{Kral}
  et~al.}{2020{\natexlab{a}}}]{Kral2020atmospheres}
{Kral}, Q; et~al., 2020{\natexlab{a}}.
\newblock Nature Astronomy, 4:769.

\bibitem[\protect\astroncite{{Kral}
  et~al.}{2020{\natexlab{b}}}]{Kral2020COsurvey}
{Kral}, Q; et~al., 2020{\natexlab{b}}.
\newblock \mnras, 497:2811.

\bibitem[\protect\astroncite{{Kral} et~al.}{2021}]{Kral2021}
{Kral}, Q; et~al., 2021.
\newblock \aap, 653:L11.

\bibitem[\protect\astroncite{{Krist} et~al.}{2012}]{Krist2012}
{Krist}, JE; et~al., 2012.
\newblock \aj, 144:45.

\bibitem[\protect\astroncite{{Krivov}}{2010}]{Krivov2010}
{Krivov}, AV, 2010.
\newblock Research in Astronomy and Astrophysics, 10:383.

\bibitem[\protect\astroncite{{Krivov} et~al.}{2006}]{Krivov2006}
{Krivov}, AV; et~al., 2006.
\newblock \aap, 455:509.

\bibitem[\protect\astroncite{{Krivov} \& {Booth}}{2018}]{Krivov2018stirring}
{Krivov}, AV et~al., 2018.
\newblock \mnras, 479:3300.

\bibitem[\protect\astroncite{{Krivov} \& {Wyatt}}{2021}]{Krivov2021}
{Krivov}, AV et~al., 2021.
\newblock \mnras, 500:718.

\bibitem[\protect\astroncite{{Lagrange} et~al.}{1987}]{Lagrange1987}
{Lagrange}, AM; et~al., 1987.
\newblock \aap, 173:289.

\bibitem[\protect\astroncite{{Lagrange} et~al.}{2009}]{Lagrange2009}
{Lagrange}, AM; et~al., 2009.
\newblock \aap, 495:335.

\bibitem[\protect\astroncite{{Li} \& {Youdin}}{2021}]{Li2021}
{Li}, R et~al., 2021.
\newblock \apj, 919:107.

\bibitem[\protect\astroncite{{Lisse} et~al.}{2009}]{Lisse2009}
{Lisse}, CM; et~al., 2009.
\newblock \apj, 701:2019.

\bibitem[\protect\astroncite{{L{\"o}hne}}{2020}]{Lohne2020}
{L{\"o}hne}, T, 2020.
\newblock \aap, 641:A75.

\bibitem[\protect\astroncite{{L{\"o}hne} et~al.}{2008}]{Lohne2008}
{L{\"o}hne}, T; et~al., 2008.
\newblock \apj, 673:1123-1137.

\bibitem[\protect\astroncite{{Long} et~al.}{2018}]{Long2018}
{Long}, F; et~al., 2018.
\newblock \apj, 869:17.

\bibitem[\protect\astroncite{{Lovell}
  et~al.}{2021{\natexlab{a}}}]{Lovell2021classIII}
{Lovell}, JB; et~al., 2021{\natexlab{a}}.
\newblock \mnras, 500:4878.

\bibitem[\protect\astroncite{{Lovell}
  et~al.}{2021{\natexlab{b}}}]{Lovell2021q1eri}
{Lovell}, JB; et~al., 2021{\natexlab{b}}.
\newblock \mnras, 506:1978.

\bibitem[\protect\astroncite{{Luppe} et~al.}{2020}]{Luppe2020}
{Luppe}, P; et~al., 2020.
\newblock \mnras, 499:3932.

\bibitem[\protect\astroncite{{Lyra} \& {Kuchner}}{2013}]{Lyra2013photoelectric}
{Lyra}, W et~al., 2013.
\newblock \nat, 499:184.

\bibitem[\protect\astroncite{{MacGregor} et~al.}{2016}]{MacGregor2016}
{MacGregor}, MA; et~al., 2016.
\newblock \apj, 823:79.

\bibitem[\protect\astroncite{{MacGregor} et~al.}{2017}]{MacGregor2017}
{MacGregor}, MA; et~al., 2017.
\newblock \apj, 842:8.

\bibitem[\protect\astroncite{{MacGregor} et~al.}{2018}]{MacGregor2018}
{MacGregor}, MA; et~al., 2018.
\newblock \apj, 869:75.

\bibitem[\protect\astroncite{{Marboeuf} et~al.}{2016}]{Marboeuf2016}
{Marboeuf}, U; et~al., 2016.
\newblock \planss, 133:47.

\bibitem[\protect\astroncite{{Marino}}{2021}]{Marino2021}
{Marino}, S, 2021.
\newblock \mnras, 503:5100.

\bibitem[\protect\astroncite{{Marino} et~al.}{2016}]{Marino2016}
{Marino}, S; et~al., 2016.
\newblock \mnras, 460:2933.

\bibitem[\protect\astroncite{{Marino}
  et~al.}{2017{\natexlab{a}}}]{Marino2017etacorvi}
{Marino}, S; et~al., 2017{\natexlab{a}}.
\newblock \mnras, 465:2595.

\bibitem[\protect\astroncite{{Marino}
  et~al.}{2017{\natexlab{b}}}]{Marino201761vir}
{Marino}, S; et~al., 2017{\natexlab{b}}.
\newblock \mnras, 469:3518.

\bibitem[\protect\astroncite{{Marino}
  et~al.}{2018{\natexlab{a}}}]{Marino2018hd107}
{Marino}, S; et~al., 2018{\natexlab{a}}.
\newblock \mnras, 479:5423.

\bibitem[\protect\astroncite{{Marino}
  et~al.}{2018{\natexlab{b}}}]{Marino2018scat}
{Marino}, S; et~al., 2018{\natexlab{b}}.
\newblock \mnras, 479:1651.

\bibitem[\protect\astroncite{{Marino} et~al.}{2019}]{Marino2019}
{Marino}, S; et~al., 2019.
\newblock \mnras, 484:1257.

\bibitem[\protect\astroncite{{Marino}
  et~al.}{2020{\natexlab{a}}}]{Marino2020hd206}
{Marino}, S; et~al., 2020{\natexlab{a}}.
\newblock \mnras, 498:1319.

\bibitem[\protect\astroncite{{Marino}
  et~al.}{2020{\natexlab{b}}}]{Marino2020gas}
{Marino}, S; et~al., 2020{\natexlab{b}}.
\newblock \mnras, 492:4409.

\bibitem[\protect\astroncite{{Marois} et~al.}{2010}]{Marois2010}
{Marois}, C; et~al., 2010.
\newblock \nat, 468:1080.

\bibitem[\protect\astroncite{{Marshall} et~al.}{2018}]{Marshall2018}
{Marshall}, JP; et~al., 2018.
\newblock \apj, 869:10.

\bibitem[\protect\astroncite{{Matr{\`a}} et~al.}{2017}]{Matra2017fomalhaut}
{Matr{\`a}}, L; et~al., 2017.
\newblock \apj, 842:9.

\bibitem[\protect\astroncite{{Matr{\`a}} et~al.}{2018}]{Matra2018mmlaw}
{Matr{\`a}}, L; et~al., 2018.
\newblock \apj, 859:72.

\bibitem[\protect\astroncite{{Matr{\`a}}
  et~al.}{2019{\natexlab{a}}}]{Matra2019betapic}
{Matr{\`a}}, L; et~al., 2019{\natexlab{a}}.
\newblock \aj, 157:135.

\bibitem[\protect\astroncite{{Matr{\`a}}
  et~al.}{2019{\natexlab{b}}}]{Matra2019twa7}
{Matr{\`a}}, L; et~al., 2019{\natexlab{b}}.
\newblock \aj, 157:117.

\bibitem[\protect\astroncite{{Matr{\`a}} et~al.}{2020}]{Matra2020}
{Matr{\`a}}, L; et~al., 2020.
\newblock \apj, 898:146.

\bibitem[\protect\astroncite{{Matthews} et~al.}{2014}]{Matthews2014pp6}
{Matthews}, BC; et~al., 2014.
\newblock Protostars and Planets VI:521.

\bibitem[\protect\astroncite{{Meng} et~al.}{2014}]{Meng2014}
{Meng}, HYA; et~al., 2014.
\newblock Science, 345:1032.

\bibitem[\protect\astroncite{{Meng} et~al.}{2015}]{Meng2015}
{Meng}, HYA; et~al., 2015.
\newblock \apj, 805:77.

\bibitem[\protect\astroncite{{Meshkat} et~al.}{2017}]{Meshkat2017}
{Meshkat}, T; et~al., 2017.
\newblock \aj, 154:245.

\bibitem[\protect\astroncite{{Michel} et~al.}{2021}]{Michel2021}
{Michel}, A; et~al., 2021.
\newblock arXiv e-prints:arXiv:2104.05894.

\bibitem[\protect\astroncite{{Miller} et~al.}{2021}]{Miller2021}
{Miller}, E; et~al., 2021.
\newblock \mnras.

\bibitem[\protect\astroncite{{Milli} et~al.}{2017}]{Milli2017}
{Milli}, J; et~al., 2017.
\newblock \aap, 599:A108.

\bibitem[\protect\astroncite{{Milli} et~al.}{2019}]{Milli2019}
{Milli}, J; et~al., 2019.
\newblock \aap, 626:A54.

\bibitem[\protect\astroncite{{Mo{\'o}r} et~al.}{2017}]{Moor2017}
{Mo{\'o}r}, A; et~al., 2017.
\newblock \apj, 849:123.

\bibitem[\protect\astroncite{{Mo{\'o}r} et~al.}{2021}]{Moor2021}
{Mo{\'o}r}, A; et~al., 2021.
\newblock \apj, 910:27.

\bibitem[\protect\astroncite{{Morales} et~al.}{2009}]{Morales2009}
{Morales}, FY; et~al., 2009.
\newblock \apj, 699:1067.

\bibitem[\protect\astroncite{{Morrison} \& {Malhotra}}{2015}]{Morrison2015}
{Morrison}, S et~al., 2015.
\newblock \apj, 799:41.

\bibitem[\protect\astroncite{{Mouillet} et~al.}{1997}]{Mouillet1997}
{Mouillet}, D; et~al., 1997.
\newblock \mnras, 292:896.

\bibitem[\protect\astroncite{{Mustill} \& {Wyatt}}{2009}]{Mustill2009}
{Mustill}, AJ et~al., 2009.
\newblock \mnras, 399:1403.

\bibitem[\protect\astroncite{{Nederlander} et~al.}{2021}]{Nederlander2021}
{Nederlander}, A; et~al., 2021.
\newblock \apj, 917:5.

\bibitem[\protect\astroncite{{Nesvold} et~al.}{2013}]{Nesvold2013}
{Nesvold}, ER; et~al., 2013.
\newblock \apj, 777:144.

\bibitem[\protect\astroncite{{Nesvold} \& {Kuchner}}{2015}]{Nesvold2015betapic}
{Nesvold}, ER et~al., 2015.
\newblock \apj, 815:61.

\bibitem[\protect\astroncite{{Nesvorn{\'y}}}{2015}]{Nesvorny2015}
{Nesvorn{\'y}}, D, 2015.
\newblock \aj, 150:73.

\bibitem[\protect\astroncite{{Norfolk} et~al.}{2021}]{Norfolk2021}
{Norfolk}, BJ; et~al., 2021.
\newblock \mnras.

\bibitem[\protect\astroncite{{{\"O}berg} et~al.}{2011}]{Oberg2011}
{{\"O}berg}, KI; et~al., 2011.
\newblock \apj, 743:L16.

\bibitem[\protect\astroncite{{Olofsson} et~al.}{2012}]{Olofsson2012}
{Olofsson}, J; et~al., 2012.
\newblock \aap, 542:A90.

\bibitem[\protect\astroncite{{Pawellek} et~al.}{2014}]{Pawellek2014}
{Pawellek}, N; et~al., 2014.
\newblock \apj, 792:65.

\bibitem[\protect\astroncite{{Pawellek} et~al.}{2019}]{Pawellek201949ceti}
{Pawellek}, N; et~al., 2019.
\newblock \mnras, 488:3507.

\bibitem[\protect\astroncite{{Pawellek} et~al.}{2021}]{Pawellek2021}
{Pawellek}, N; et~al., 2021.
\newblock \mnras.

\bibitem[\protect\astroncite{{Pearce} \& {Wyatt}}{2014}]{Pearce2014}
{Pearce}, TD et~al., 2014.
\newblock \mnras, 443:2541.

\bibitem[\protect\astroncite{{Pearce} \& {Wyatt}}{2015}]{Pearce2015doublering}
{Pearce}, TD et~al., 2015.
\newblock \mnras, 453:3329.

\bibitem[\protect\astroncite{{Pearce} et~al.}{2020}]{Pearce2020}
{Pearce}, TD; et~al., 2020.
\newblock \mnras, 498:2798.

\bibitem[\protect\astroncite{{Perrin} et~al.}{2015}]{Perrin2015}
{Perrin}, MD; et~al., 2015.
\newblock \apj, 799:182.

\bibitem[\protect\astroncite{{Perrot} et~al.}{2016}]{Perrot2016}
{Perrot}, C; et~al., 2016.
\newblock \aap, 590:L7.

\bibitem[\protect\astroncite{{Petit} et~al.}{2011}]{Petit2011}
{Petit}, JM; et~al., 2011.
\newblock \aj, 142:131.

\bibitem[\protect\astroncite{{Plavchan} et~al.}{2005}]{Plavchan2005}
{Plavchan}, P; et~al., 2005.
\newblock \apj, 631:1161.

\bibitem[\protect\astroncite{{Quillen}}{2006}]{Quillen2006}
{Quillen}, AC, 2006.
\newblock \mnras, 372:L14.

\bibitem[\protect\astroncite{{Rebollido} et~al.}{2020}]{Rebollido2020}
{Rebollido}, I; et~al., 2020.
\newblock \aap, 639:A11.

\bibitem[\protect\astroncite{{Ren} et~al.}{2021}]{Ren2021}
{Ren}, B; et~al., 2021.
\newblock \apj, 914:95.

\bibitem[\protect\astroncite{{Rhee} et~al.}{2008}]{Rhee2008}
{Rhee}, JH; et~al., 2008.
\newblock \apj, 675:777-783.

\bibitem[\protect\astroncite{{Ricci} et~al.}{2015}]{Ricci2015}
{Ricci}, L; et~al., 2015.
\newblock \apj, 798:124.

\bibitem[\protect\astroncite{{Richert} et~al.}{2018}]{Richert2017}
{Richert}, AJW; et~al., 2018.
\newblock \apj, 856:41.

\bibitem[\protect\astroncite{{Rieke} et~al.}{2021}]{Rieke2021}
{Rieke}, GH; et~al., 2021.
\newblock \apj, 918:71.

\bibitem[\protect\astroncite{{Rigley} \& {Wyatt}}{2020}]{Rigley2020}
{Rigley}, JK et~al., 2020.
\newblock \mnras, 497:1143.

\bibitem[\protect\astroncite{{Rodigas} et~al.}{2015}]{Rodigas2015}
{Rodigas}, TJ; et~al., 2015.
\newblock \apj, 798:96.

\bibitem[\protect\astroncite{{Romero} et~al.}{2021}]{Romero2021}
{Romero}, C; et~al., 2021.
\newblock \aap, 651:A34.

\bibitem[\protect\astroncite{{Rosotti} et~al.}{2020}]{Rosotti2020}
{Rosotti}, GP; et~al., 2020.
\newblock \mnras, 495:173.

\bibitem[\protect\astroncite{{Sai} et~al.}{2015}]{Sai2015}
{Sai}, S; et~al., 2015.
\newblock \pasj, 67:20.

\bibitem[\protect\astroncite{{Schneider} et~al.}{2006}]{Schneider2006}
{Schneider}, G; et~al., 2006.
\newblock \apj, 650:414.

\bibitem[\protect\astroncite{{Schneider} et~al.}{2014}]{Schneider2014}
{Schneider}, G; et~al., 2014.
\newblock \aj, 148:59.

\bibitem[\protect\astroncite{Schneiderman et~al.}{2021}]{Schneiderman2021}
Schneiderman, T; et~al., 2021.
\newblock Nature, 598:425.

\bibitem[\protect\astroncite{{Sefilian} et~al.}{2021}]{Sefilian2020}
{Sefilian}, AA; et~al., 2021.
\newblock \apj, 910:13.

\bibitem[\protect\astroncite{{Sepulveda} et~al.}{2019}]{Sepulveda2019}
{Sepulveda}, AG; et~al., 2019.
\newblock \apj, 881:84.

\bibitem[\protect\astroncite{{Sezestre} et~al.}{2017}]{Sezestre2017}
{Sezestre}, {\'E}; et~al., 2017.
\newblock \aap, 607:A65.

\bibitem[\protect\astroncite{{Sezestre} et~al.}{2019}]{Sezestre2019}
{Sezestre}, {\'E}; et~al., 2019.
\newblock \aap, 626:A2.

\bibitem[\protect\astroncite{{Sibthorpe} et~al.}{2018}]{Sibthorpe2018}
{Sibthorpe}, B; et~al., 2018.
\newblock \mnras, 475:3046.

\bibitem[\protect\astroncite{{Slettebak}}{1975}]{Slettebak1975}
{Slettebak}, A, 1975.
\newblock \apj, 197:137.

\bibitem[\protect\astroncite{{Stammler} et~al.}{2019}]{Stammler2019}
{Stammler}, SM; et~al., 2019.
\newblock \apjl, 884:L5.

\bibitem[\protect\astroncite{{Str{\o}m} et~al.}{2020}]{Strom2020}
{Str{\o}m}, PA; et~al., 2020.
\newblock \pasp, 132:101001.

\bibitem[\protect\astroncite{{Su} et~al.}{2006}]{Su2006}
{Su}, KYL; et~al., 2006.
\newblock \apj, 653:675.

\bibitem[\protect\astroncite{{Su} et~al.}{2019}]{Su2019}
{Su}, KYL; et~al., 2019.
\newblock \aj, 157:202.

\bibitem[\protect\astroncite{{Th{\'e}bault} \& {Augereau}}{2007}]{Thebault2007}
{Th{\'e}bault}, P et~al., 2007.
\newblock \aap, 472:169.

\bibitem[\protect\astroncite{{Th{\'e}bault} \& {Wu}}{2008}]{Thebault2008}
{Th{\'e}bault}, P et~al., 2008.
\newblock \aap, 481:713.

\bibitem[\protect\astroncite{{van Lieshout} et~al.}{2014}]{vanLieshout2014}
{van Lieshout}, R; et~al., 2014.
\newblock \aap, 571:A51.

\bibitem[\protect\astroncite{{Veras}}{2021}]{Veras2021}
{Veras}, D, 2021.
\newblock arXiv e-prints:arXiv:2106.06550.

\bibitem[\protect\astroncite{{Visser} et~al.}{2009}]{Visser2009}
{Visser}, R; et~al., 2009.
\newblock \aap, 503:323.

\bibitem[\protect\astroncite{{Watt} et~al.}{2021}]{Watt2021}
{Watt}, L; et~al., 2021.
\newblock \mnras, 502:2984.

\bibitem[\protect\astroncite{{Weidenschilling}}{1977}]{Weidenschilling1977drag}
{Weidenschilling}, SJ, 1977.
\newblock \mnras, 180:57.

\bibitem[\protect\astroncite{{Welsh} et~al.}{1998}]{Welsh1998}
{Welsh}, BY; et~al., 1998.
\newblock \aap, 338:674.

\bibitem[\protect\astroncite{{Whipple}}{1973}]{Whipple1973}
{Whipple}, FL, 1973.
\newblock NASA Special Publication, 319:355.

\bibitem[\protect\astroncite{{Wisdom}}{1980}]{Wisdom1980}
{Wisdom}, J, 1980.
\newblock \aj, 85:1122.

\bibitem[\protect\astroncite{{Wisniewski} et~al.}{2019}]{Wisniewski2019}
{Wisniewski}, JP; et~al., 2019.
\newblock \apjl, 883:L8.

\bibitem[\protect\astroncite{{Wyatt}}{2008}]{Wyatt2008}
{Wyatt}, MC, 2008.
\newblock \araa, 46:339.

\bibitem[\protect\astroncite{{Wyatt} et~al.}{1999}]{Wyatt1999}
{Wyatt}, MC; et~al., 1999.
\newblock \apj, 527:918.

\bibitem[\protect\astroncite{{Wyatt}
  et~al.}{2007{\natexlab{a}}}]{Wyatt2007Astars}
{Wyatt}, MC; et~al., 2007{\natexlab{a}}.
\newblock \apj, 663:365.

\bibitem[\protect\astroncite{{Wyatt}
  et~al.}{2007{\natexlab{b}}}]{Wyatt2007hotdust}
{Wyatt}, MC; et~al., 2007{\natexlab{b}}.
\newblock \apj, 658:569.

\bibitem[\protect\astroncite{{Yelverton} \& {Kennedy}}{2018}]{Yelverton2018}
{Yelverton}, B et~al., 2018.
\newblock \mnras, 479:2673.

\bibitem[\protect\astroncite{{Yelverton} et~al.}{2019}]{Yelverton2019}
{Yelverton}, B; et~al., 2019.
\newblock \mnras, 488:3588.

\bibitem[\protect\astroncite{{Yelverton} et~al.}{2020}]{Yelverton2020}
{Yelverton}, B; et~al., 2020.
\newblock \mnras, 495:1943.

\bibitem[\protect\astroncite{{Youdin} \& {Goodman}}{2005}]{Youdin2005}
{Youdin}, AN et~al., 2005.
\newblock \apj, 620:459.

\bibitem[\protect\astroncite{{Youngblood} et~al.}{2021}]{Youngblood2021}
{Youngblood}, A; et~al., 2021.
\newblock arXiv e-prints:arXiv:2108.11965.

\bibitem[\protect\astroncite{{Zsom} et~al.}{2010}]{Zsom2010}
{Zsom}, A; et~al., 2010.
\newblock \aap, 513:A57.

\bibitem[\protect\astroncite{{Zuckerman} et~al.}{1995}]{Zuckerman1995}
{Zuckerman}, B; et~al., 1995.
\newblock \nat, 373:494.

\bibitem[\protect\astroncite{{Zuckerman} \& {Song}}{2012}]{Zuckerman2012}
{Zuckerman}, B et~al., 2012.
\newblock \apj, 758:77.

\end{thebibliography}

%\blankpage
%\printindex[aindx]                 % to print author index
%\printindex                        % to print subject index

\end{document}